\documentclass{jfm}
\usepackage{graphicx}
\usepackage{multirow}
\usepackage{amsmath}
\usepackage{amssymb}
\usepackage{float}
\usepackage{epstopdf, epsfig}
\usepackage{xcolor}
\usepackage{tikz}
\usepackage{mathabx}

\definecolor{myred}{RGB}{192, 0, 0}
\definecolor{mycyan}{RGB}{0, 112, 192}
\definecolor{myblue}{RGB}{0, 0, 255}

\shorttitle{Flow oscillations in high-speed double cones}
\shortauthor{G. Kumar, V. Sasidharan, A. G. Kumara and S. Duvvuri}

\title{A model for frequency scaling of flow oscillations in high-speed double cones}

\author{Gaurav Kumar,
  Vaisakh Sasidharan, 
  Akshaya G. Kumara
 \and Subrahmanyam Duvvuri\corresp{\email{subrahmanyam@iisc.ac.in}}}

\affiliation{Turbulent Shear Flow Physics and Engineering Laboratory\\Department of Aerospace Engineering, Indian Institute of Science, Bengaluru 560 012}

\begin{document}

\maketitle

\begin{abstract}
Coherent small-amplitude unsteadiness of the shock wave and the separation region over a canonical double cone flow, termed in literature as oscillation-type unsteadiness, is experimentally studied at Mach 6. The double cone model is defined by three non-dimensional geometric parameters: fore- and aft-cone angles ($\theta_1$ and $\theta_2$), and ratio of the conical slant lengths ($\Lambda$). Previous studies of oscillations have been qualitative in nature, and mostly restricted to a special case of the cone model with fixed $\theta_1 = 0^\circ$ and $\theta_2 = 90^\circ$ (referred to as the spike-cylinder model), where $\Lambda$ becomes the sole governing parameter. In the present effort we investigate the self-sustained flow oscillations in the $\theta_1$-$\Lambda$ parameter space for fixed $\theta_2 = 90^\circ$ using time-resolved schlieren visualization. The experiments reveal two distinct sub-types of oscillations, characterized by the motion (or lack thereof) of the separation point on the fore-cone surface. The global time scale associated with flow oscillation is extracted using spectral proper orthogonal decomposition. The non-dimensional frequency (Strouhal number) of oscillation is seen to exhibit distinct scaling for the two oscillation sub-types. The relationship observed between the local flow properties, instability of the shear layer, and geometric constraints on the flow suggests that an aeroacoustic feedback mechanism sustains the oscillations. Based on this insight, a simple model with no empiricism is developed for the Strouhal number. The model predictions are found to match well with experimental measurements. The model provides helpful physical insight into the nature of the self-sustained flow oscillations over a double cone at high-speeds.
\end{abstract}


\section{Introduction}\label{sec:Introduction}
\begin{figure}
	\centering
	\includegraphics[width=0.7\linewidth]{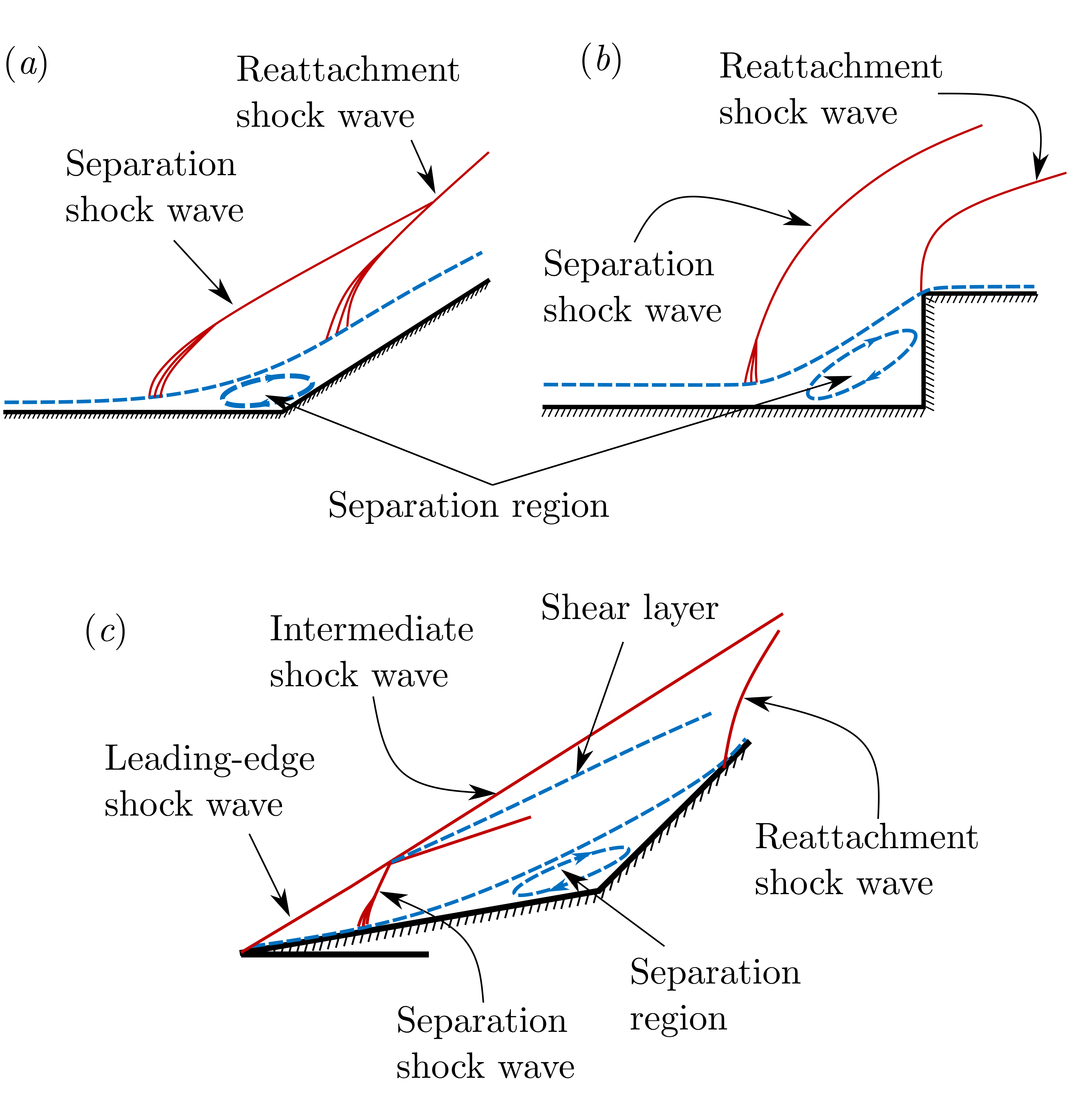}
	\caption{A schematic illustrations of different flows with shock wave/boundary layer interaction (SBLI) and shock wave/separation region interaction (SSRI). (\emph{a} and \emph{b}) Formation of separation and reattachment shock waves when a supersonic flow is turned due to an inclined ramp or a forward facing step, respectively. (\emph{c}) Interaction between the leading-edge shock wave and the separation region for a double ramp. Intersection of the separation shock wave with the leading-edge shock wave generates an intermediate shock wave and a shear layer, both of which are in the vicinity of the shear layer that envelopes the separation bubble.}
	\label{fig:Intro}
\end{figure}

High-speed flow (supersonic or hypersonic flow) over a surface with a compression corner typically features a separated boundary layer and associated shock waves, as illustrated in figures \ref{fig:Intro} (\emph{a}) and \ref{fig:Intro} (\emph{b}). The separation and reattachment shock waves can exhibit low-frequency oscillatory motion -- a phenomenon that has extensively been studied under the class of problems referred to as shock wave/boundary layer interaction (SBLI) \citep[for instance, see][and references therein]{beresh2002relationship,touber2011low,clemens2014low, gaitonde2015sbli, murugan2016shock,estruch2018separated, gaitonde2023sbli}. SBLI studies generally consider a sufficiently developed boundary layer where the leading edge of the wall has little influence on the local flow dynamics in the SBLI region. In contrast, an SBLI type scenario situated close to the leading edge of the wall exhibits distinct and complex flow features arising from interactions between the leading edge, separated and reattachment shock waves, and separated boundary/shear layer, as illustrated in figure \ref{fig:Intro} (\emph{c}). The intersection of the leading-edge shock wave with the unsteady separation shock wave creates an intermediate shock wave. This shock wave systems sets the flow conditions over the separation region, and viscous interactions in the separation region can lead to large-spatial-amplitude and low-frequency oscillatory motion of the entire shock-wave system. In one class of such unsteady flows, the shear layer (\emph{i.e.}, the separated boundary layer) that forms over the separation region/bubble is susceptible to flow instabilities, and growth of disturbances arising from such instabilities leads to interesting unsteady coupled dynamics of the overall flow system. These types of unsteady flows are distinct from SBLI and are classified as shock wave/separation region interaction (SSRI) problems \citep{duvvuri2023shock}.

\begin{figure}
	\centering
	\includegraphics[width=0.45\linewidth]{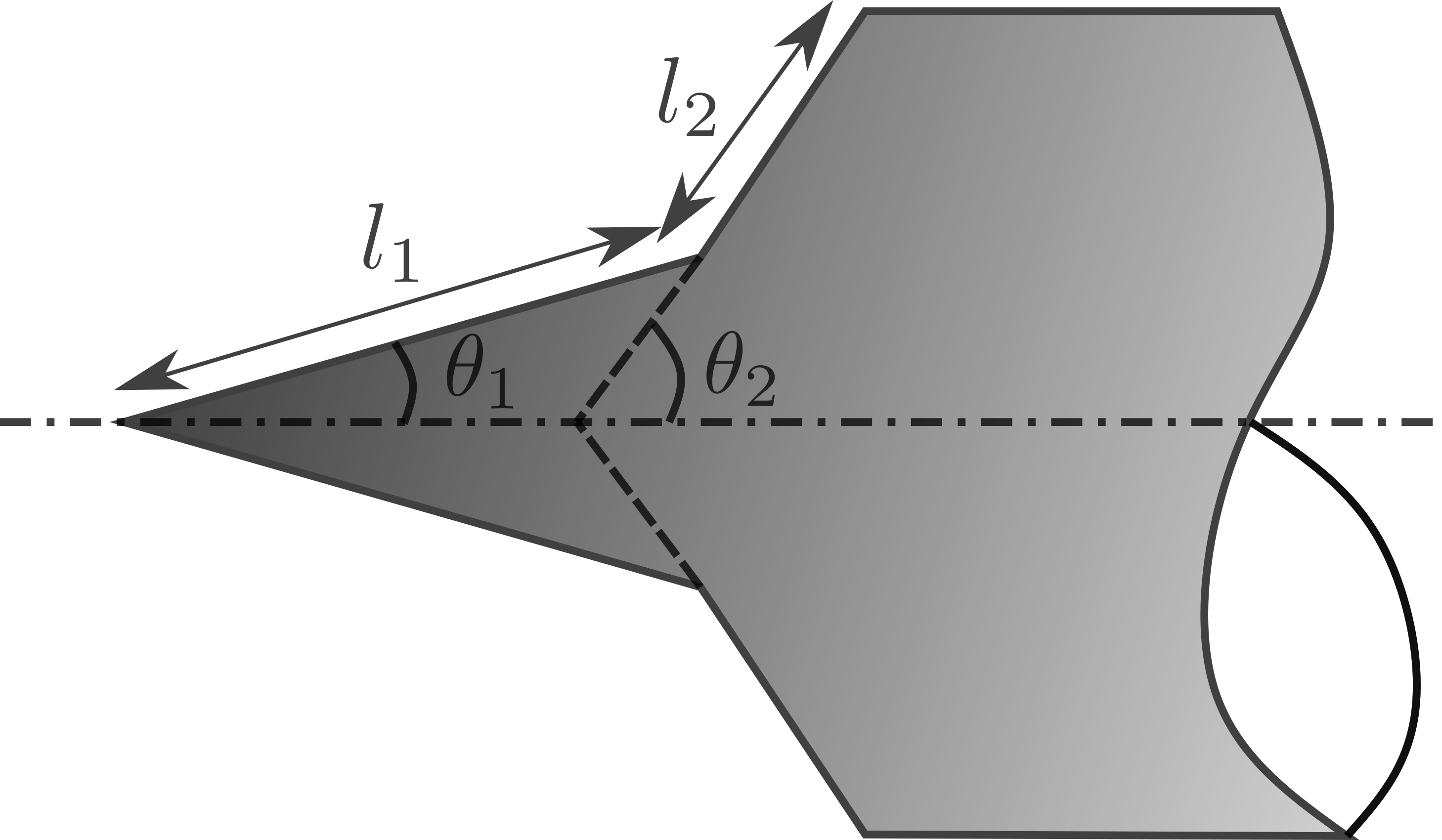}
	\caption{A schematic of the double-cone model. Fore-cone half-angle $\theta_1$, aft-cone half-angle $\theta_2$, and slant length ratio $\Lambda = l_2/l_1$ are the three non-dimensional geometric parameters that completely define the model.}
	\label{fig:doubleCone}
\end{figure}

High-speed flow over a double cone generates unsteadiness due to SSRI, and makes for a good canonical problem for a detailed study of the flow dynamics associated with SSRI. The double cone model comprises of two adjacent right circular conical sections along a single axis (see figure~\ref{fig:doubleCone}). The model geometry is described by the two cone half-angles $\theta_1$, $\theta_2$ and slant lengths $l_1$, $l_2$ of the conical sections.  It is noted that the double wedge geometry is the 2D version of the double cone, where the conical sections are replaced by rectangular wedges of half-angles $\theta_1$, $\theta_2$ and slant lengths $l_1$, $l_2$). Both double cone and double wedge configurations can be fully defined by three non-dimensional parameters, taken here to be $\theta_1$, $\theta_2$ and $\Lambda = l_2/l_1$.

\begin{figure}
	\centering
	\includegraphics[width=\linewidth]{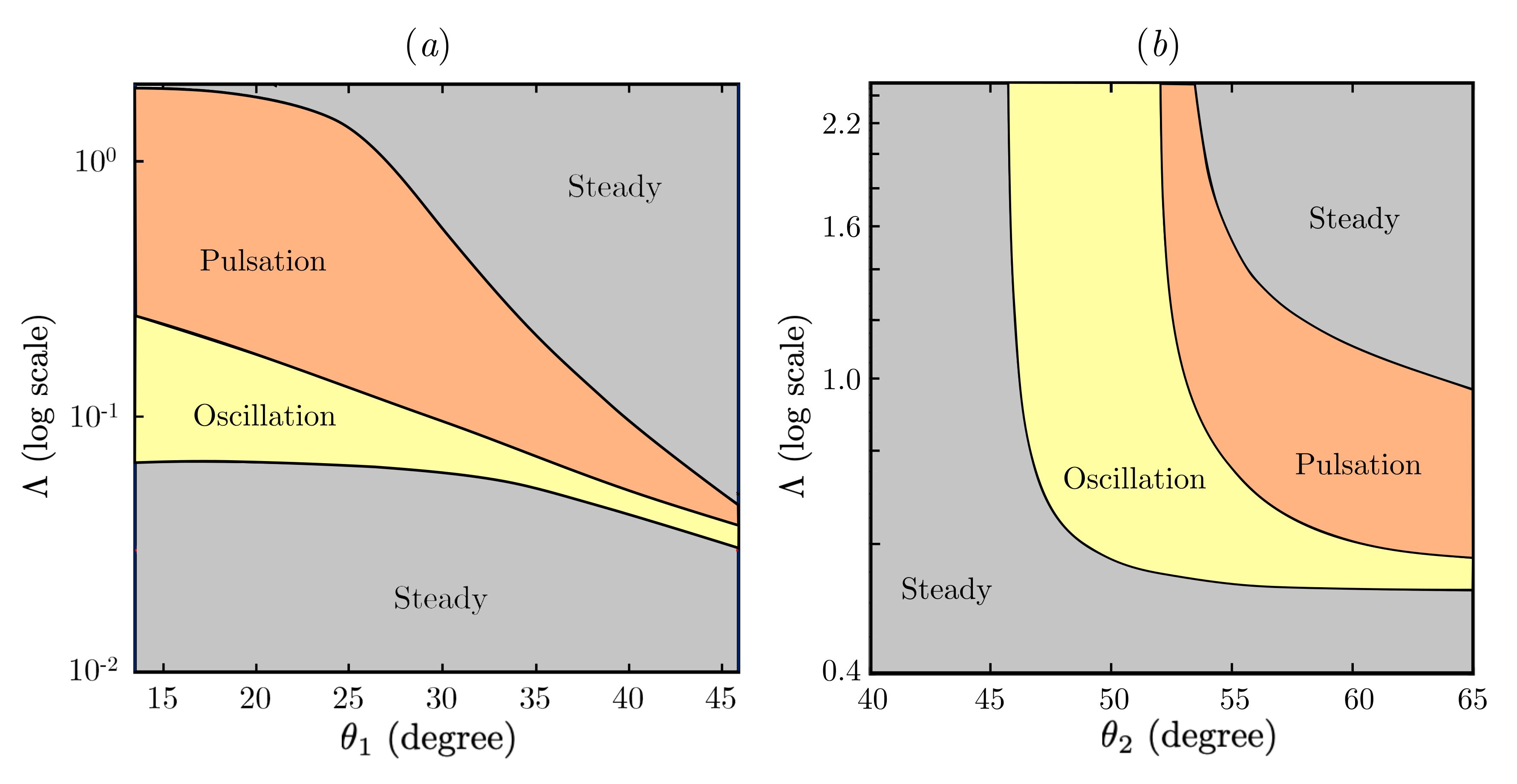}
	\caption{(\emph{a}) $\theta_1$-$\Lambda$ parameter space for a double cone of $\theta_2 = 90^\circ$ in a Mach 6 flow, with boundaries between flow states generated empirically from experimental results of \cite{SD2021JFM}. (\emph{b}) $\theta_2$-$\Lambda$ parameter for a double wedge of $\theta_1 = 30^\circ$ in a Mach 7 flow, with boundaries between flow states generated empirically from computational results of \cite{kumar2021modes}.}
	\label{fig:subclasses_map}
\end{figure}


A special case of the general double cone configuration is the spike-cylinder model, where $\theta_1$ and $\theta_2$ are fixed at $0^\circ$ and $90^\circ$, respectively. The spike-cylinder geometry is defined by a single non-dimensional parameter, \emph{i.e.}, $\Lambda$. High-speed spike-cylinder flow has attracted considerable research attention since the 1950s \citep[see][and references therein]{sahoo2021spikec, duvvuri2023shock}. Early experimental studies of the flow clearly revealed unsteady and periodic flow phenomena \citep{maull1960hypersonic, wood1962hypersonic, kenworthy1975study, kenworthy1978study}. As noted above, the spatial extent of unsteady motions in the flow (for a certain range of $\Lambda$) can be much larger than unsteadiness observed in SBLI scenarios. \cite{kenworthy1975study} identified two distinct states of unsteadiness and labeled them as ``pulsation'' and ``oscillation''. Flow pulsation is characterized by large-scale unsteady motion of shock waves driven by a periodic growth and collapse of the separation bubble. Whereas flow oscillation is a distinct, and relatively smaller scale, unsteadiness of shock waves which is driven by flow disturbances in the shear layer that forms over the separation bubble. For a given flow Reynolds and Mach numbers, the state of flow unsteadiness is determined by $\Lambda$. See \cite{duvvuri2023shock} for a more detailed discussion on the unsteady flow states of pulsation and oscillation.

In recent years there has been an interest in understanding the double cone/ramp flow behavior with variation in more than one geometric parameter. \cite{SD2021JFM} experimentally studied the double cone flow in the $\theta_1$-$\Lambda$ parameter space with $\theta_2$ fixed at $90^\circ$ (cone-cylinder model), while \cite{kumar2021modes} studied the double wedge flow through computations in the $\theta_2$-$\Lambda$ parameter space with $\theta_1$ fixed at $30^\circ$. \cite{hornung2021unsteadiness} performed computations of the double cone flow with variation in all three geometric parameters. Results from these studies show that even as flow moves into a higher dimensional geometric parameter space from the singe-parameter spike-cylinder geometry, it only exhibits two distinct states of unsteadiness: pulsation and oscillation. These unsteady flow states are qualitatively the same as the pulsation and oscillation states reported in earlier literature on spike-cylinder flow, and hence the same terminology was used by \cite{SD2021JFM}, \cite{kumar2021modes}, and \cite{hornung2021unsteadiness} to refer to the two unsteady flow states. These studies also reveal that the boundaries of the unsteady flow states in the $\theta_1$-$\theta_2$-$\Lambda$ are non-linear (see figure~\ref{fig:subclasses_map}).  

\begin{figure}
	\centering
	\includegraphics[width=\linewidth]{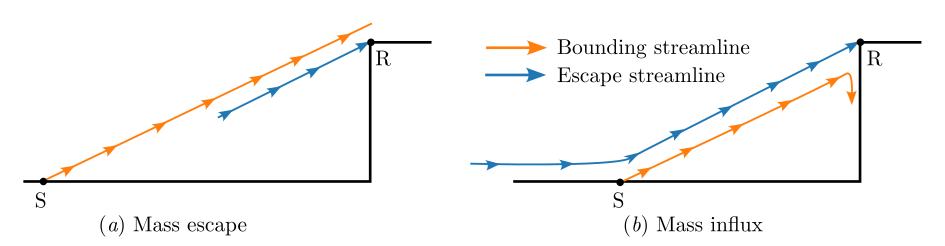}
	\caption{``Bounding streamline'' and ``escape streamline'' shown by \cite{feszty2004driving} for oscillation type flow unsteadiness over a spike-cylinder model. Here ‘S’ and ‘R’ represent the separation and reattachment points of the shear layer, respectively. This figure is adapted from \cite{feszty2004driving}}.
	\label{fig:ELSH}
\end{figure}

The present effort is aimed at obtaining a detailed understanding of the oscillation state of unsteadiness in the high-speed double cone flow. The current understanding of oscillations derived from available literature is qualitative in nature, mostly based on the hypothesis of \cite{kenworthy1978study}. Based on an experimental study of the spike-cylinder flow, \cite{kenworthy1978study} concludes that oscillations are sustained through viscous instabilities of the shear layer, and that oscillations originate from an imbalance between two pressure quantities -- the so-called ``reattachment pressure'' generated by the shear layer and the ``reattachment pressure'' required by geometric constraints. This hypothesis led to an ``energetic shear-layer'' model \citep{kenworthy1978study} which essentially tries to explain the periodic expansion and contraction of the separation region (which are observed in oscillation-type unsteadiness) based on possible mass exchange near the reattachment point, as illustrated in figure~\ref{fig:ELSH}. Fluid mass is thought to ``escape'' from the separation bubble when the ``bounding streamline'' is above the ``escape streamline'' (see figure~\ref{fig:ELSH} (\emph{a})), and fluid mass is thought to be injected into the separation bubble when the ``bounding streamline'' is below the ``escape streamline'' (see figure~\ref{fig:ELSH} (\emph{b})). \cite{feszty2004driving} supports this hypothesis on the basis of a computational study. While the mass exchange mechanism outlined here provides a physical picture of flow oscillations, the insights obtained from the same are not directly amenable to quantitative modeling.

The present study of oscillation dynamics in a high-speed double-cone flow takes an aeroacoustic viewpoint. This is partly motivated by broad similarities between double cone flow oscillations and periodic unsteadiness in canonical open cavity flows (see \S\ref{sec:mechanism}). The compressible free-shear-layer and the large separated flow region in an open cavity flow exhibit low-frequency unsteadiness \citep{rossiter1964wind}. The specific aims of the present study are twofold:
\begin{enumerate}
    \item obtain a quantitative understanding of the dependence of spatio-temporal scales of oscillations on the governing geometric parameters;
    \item develop a model to predict the Strouhal number (non-dimensional frequency) of flow oscillations.
\end{enumerate}
To achieve these aims, an extensive set of double cone flow experiments were performed at Mach 6. In these experiments $\theta_1$ and $\Lambda$ were varied while $\theta_2$ was fixed at $90^\circ$, and the experiments were focused on the oscillation region of the $\theta_1$-$\Lambda$ parameter space. The experimental setup is briefly described in \S2 and experimental results are presented in \S3. One of the important revelations from these experiments is the presence of two distinct sub-types of oscillations in the double cone flow. \S3 provides a detailed discussion of the two oscillation sub-types, which are labeled here as ``free-oscillations'' and ``anchored-oscillations.'' The aeroacoustic model is described and results from the same are presented in \S4. Finally, a set of brief concluding remarks are presented in \S5.

\section{Experimental setup}\label{sec:Experimental_setup}

\begin{figure}
	\centering
	\includegraphics[width=\linewidth]{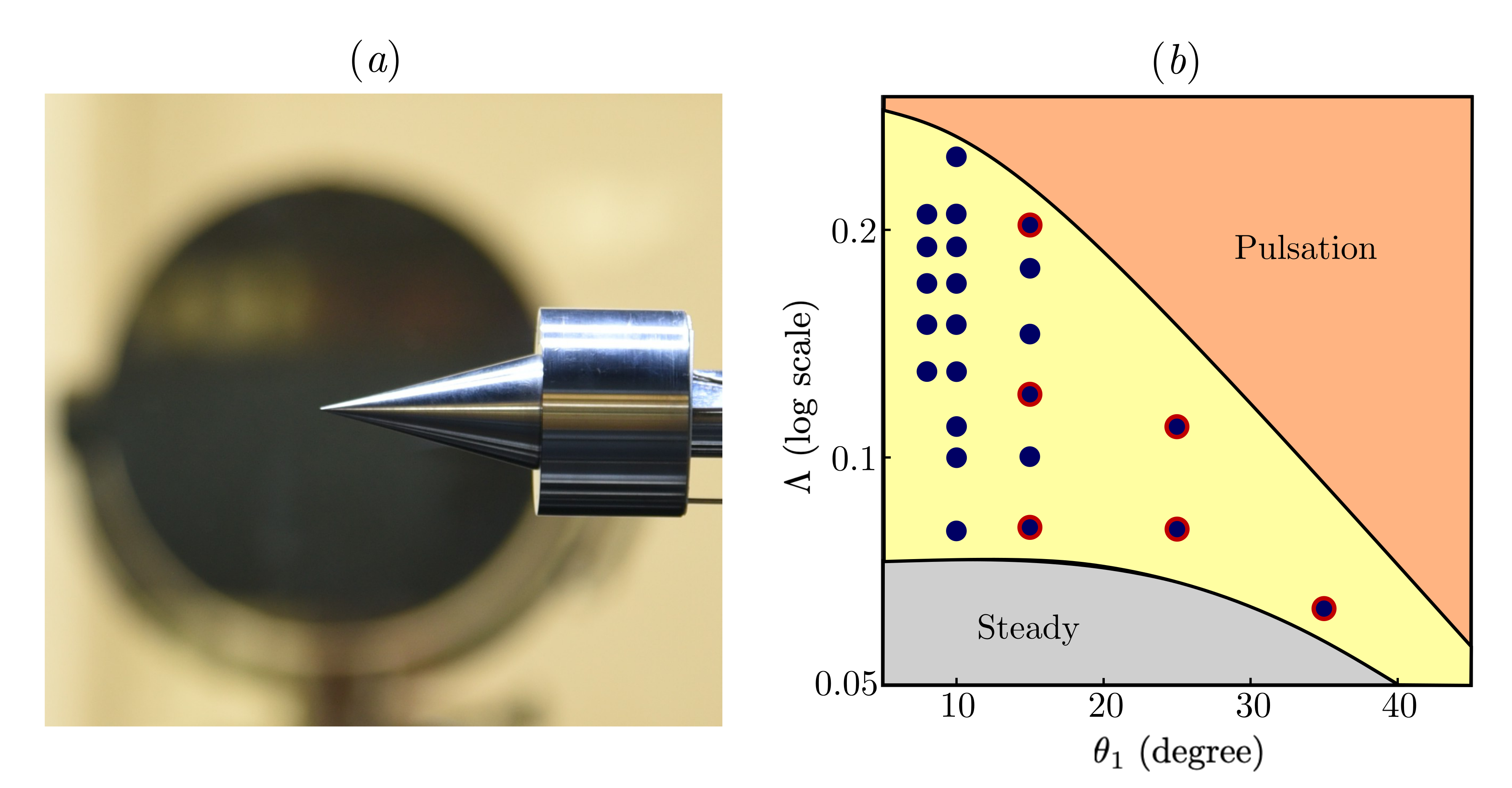}
	\caption{(\emph{a}) A double cone model installed in the test section of the Roddam Narasimha Hypersonic Wind Tunnel. (\emph{b}) Data markers (filled circles in dark blue) denote individual experiments; in each experiment data was obtained for a specific combination of $\theta_1$-$\Lambda$ while $\theta_2$ remains fixed at $90^\circ$ across all experiments. Data markers with a red outline denote experiments performed by \cite{SD2021JFM}.}
	\label{fig:mounted_model}
\end{figure}

Experiments were conducted in the Roddam Narasimha Hypersonic Wind Tunnel (RNHWT) at the Indian Institute of Science. RNHWT is a 0.5 meter diameter enclosed free-jet facility (pressure-vacuum type) capable of producing free-stream flow in the Mach number range 6 to 10 with dry air as the working fluid. All the experiments for the present study were carried out at free-stream Mach number $M_\infty = 6$ with stagnation temperature $T_0$ and pressure $P_0$ set to 455 K and 11.1 bar, respectively. The corresponding free-stream unit Reynolds number $Re_\infty = 10^7$ m$^{-1}$, based on density $\rho_\infty$, speed $U_\infty$, and dynamic viscosity $\mu_\infty$ defined in the free-stream. All the double cone models used in the present study have an aft-cone angle $\theta_2 = 90^\circ$, and the test cases are identified by the fore-cone angle $\theta_1$ and cone slant length ratio $\Lambda$. Figure~\ref{fig:mounted_model} (\emph{a}) shows a representative image of a double-cone model installed in the RNHWT test section. The value of $\theta_1$ was varied between $8^\circ$ and $35^\circ$ across experiments, and for any fixed value of $\theta_1$, the $\Lambda$ values were selected such that an oscillation-type unsteady flow state is realized. Figure~\ref{fig:mounted_model} (\emph{b}) provides an overview of all the different $\theta_1$-$\Lambda$ combinations for which experimental data was obtained. It is noted that each data marker in figure~\ref{fig:mounted_model} (\emph{b}) denotes an individual experiment. In addition to data obtained from the present set of experiments, data from the earlier study of \cite{SD2021JFM} is also used here for analysis and modeling (in \S3 and \S4). Combinations of $\theta_1$-$\Lambda$ for which data is borrowed from \cite{SD2021JFM} are identified in figure~\ref{fig:mounted_model} (\emph{b}).

In all experiments, unsteady motions in the flow were visualized by employing a time-resolved schlieren technique. A high-power pulsed diode laser (Cavilux Smart, 640 nm wavelength, 10 ns pulse width) was used as the light source and a high-speed camera (Phantom V1612) was used for imaging. Schlieren images were recorded at frame rates ranging between 56000 and 140000 frames-per-second. The number of oscillation cycles captured by the schlieren data range between 400 and 1200 across all experiments. These sampling parameters provide a temporal resolution in the range 20 to 56 images per oscillation cycle. Overall, the data sets obtained in this study contain good spatio-temporal resolution and are sufficiently long (data length) to allow for detailed analysis. Given the symmetry of double cone about the model axis, coverage area of schlieren images was restricted to the top half of the model.

\section{Experimental results and analysis}\label{sec:Results}

\begin{figure}
	\centering
	\includegraphics[width=\linewidth,trim=0 0 0 0, clip=true]{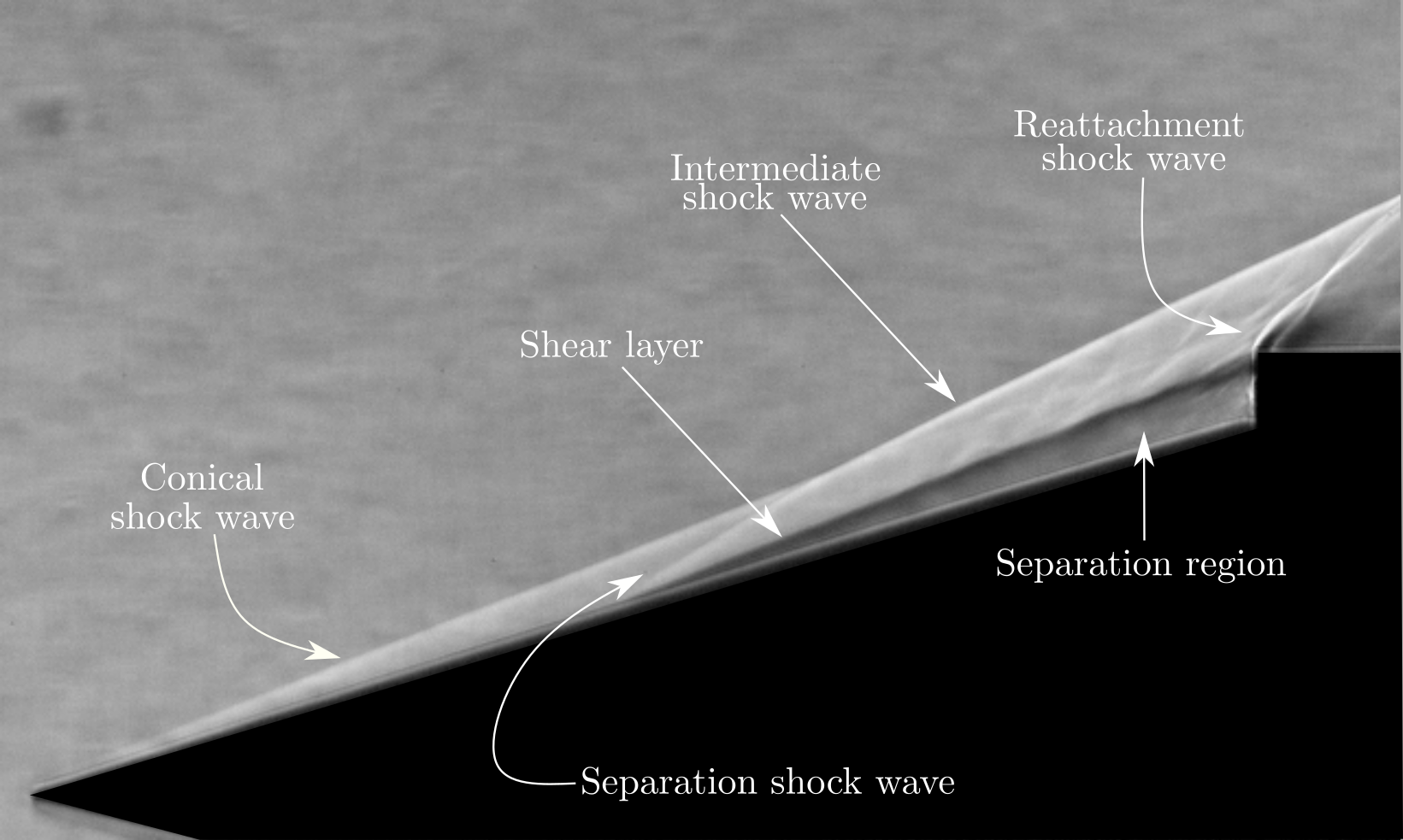}
	\caption{An instantaneous schlieren snapshot of the flow for [$\theta_1,\Lambda$] = [$15^\circ$,0.06]. Due to the relatively small value of $\Lambda$, this flow is nominally steady (see video in supplementary material).}
	\label{fig:steadyFlow}
\end{figure}
\begin{figure}
	\centering
	\includegraphics[width=0.9\linewidth]{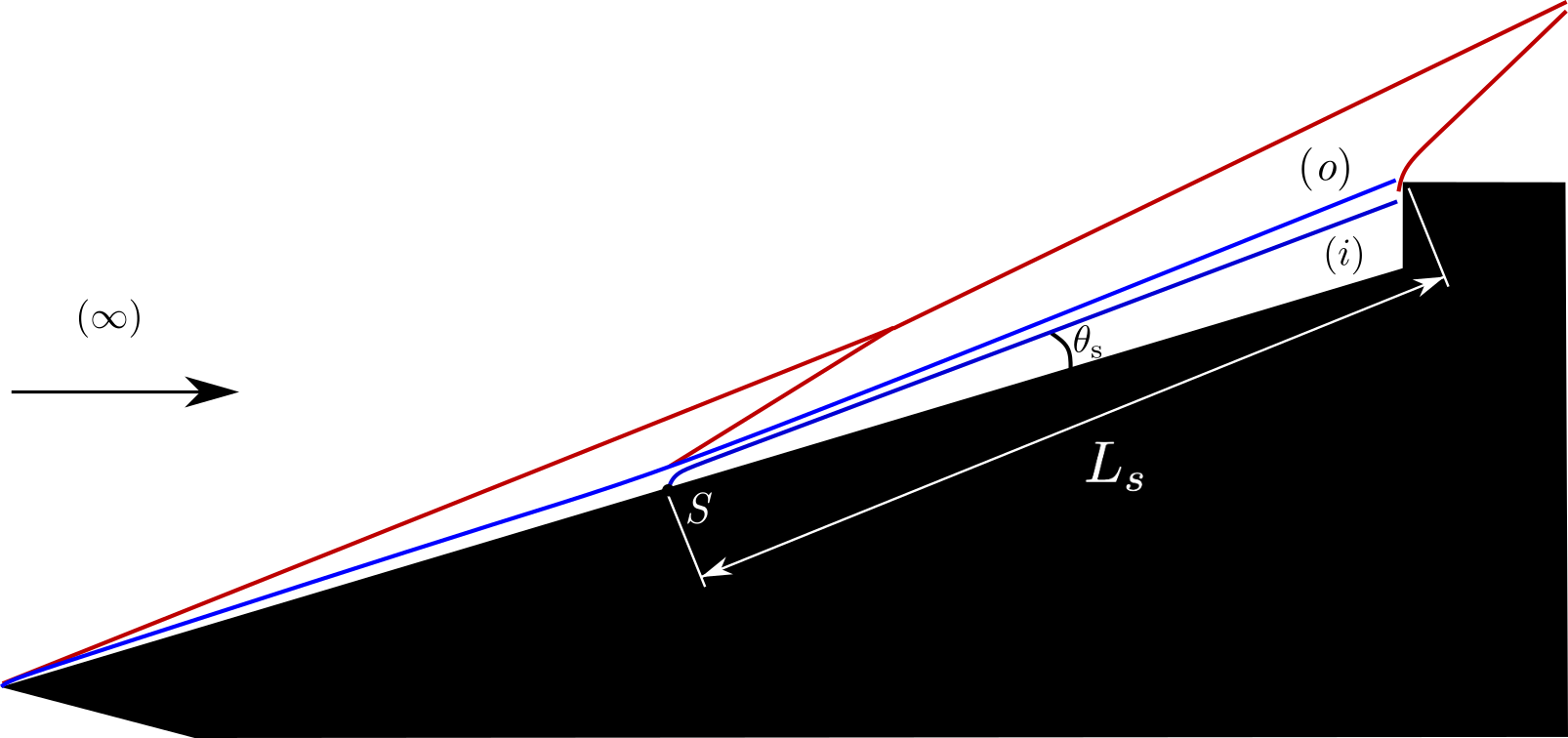}
	\caption{A schematic representation of the flow shown in figure~\ref{fig:steadyFlow}. Shock waves are represented in red and shear layer in blue. $(\infty)$ represent the incoming free-stream flow. Location $S$ marks the separation point, and flow regions inside and outside separation bubble are marked by ($i$) and ($o$). The length of the shear layer, measured along the shear layer between the separation and the reattachment points, is denoted by $L_s$. $\theta_s$ represents the angle formed by the shear layer with the fore-cone surface.}
	\label{fig:shearLayerScheme}
\end{figure}

It is instructive to begin the discussion on oscillations by considering the flow structure in a scenario where $\Lambda \ll 1$, and thereby the flow is nominally steady. Figure~\ref{fig:steadyFlow} shows a representative example of such a flow with [$\theta_1,\Lambda$] = [$15^\circ$,0.06]. It is noted that this particular combination of $\theta_1$ and $\Lambda$ falls in the steady region of the parameter space shown in figure~\ref{fig:mounted_model} (\emph{b}); it is not represented by a data marker in the figure since it falls outside the oscillations flow regime (which is of present interest). From the schlieren image in figure~\ref{fig:steadyFlow}, it is clearly seen that an attached conical shock wave at the nose and a strong reattachment shock wave above the cylinder shoulder are formed. This is consistent with expectations since $\theta_1$ is less than $\theta_d$, and $\theta_2$ is greater than $\theta_d$; here $\theta_d = 55.5^\circ$ is the conical shock wave detachment angle corresponding to $M_{\infty} = 6$. Flow turning by the aft-cone results in boundary layer flow separation over the fore-cone surface and generation of a separation shock wave. The intersection of separation and conical shock waves give rise to the intermediate shock wave, which then interacts with the reattachment shock wave at a downstream location. A compressible shear layer forms over the separated flow region, and it reattaches at the aft-cone (cylinder) shoulder. The shear layer is a region of large density gradient in the cross-stream direction, and hence it is easily identified in the schlieren image. All these flow features are schematically illustrated in figure~\ref{fig:shearLayerScheme}. Point $S$ in the figure marks the separation point, and $L_s$, $\theta_s$ represent the length of the shear layer and the shear layer angle with the cone surface, respectively. The markings `$i$' and `$o$' in the figure denote the flow regions inside and outside of the separation bubble, and they are used as subscripts to identify physical quantities in corresponding regions of the flow. It is noted that the flow for this geometric configuration is taken to be nominally steady, notwithstanding some non-periodic jitters of very small amplitude arising from incipient instabilities of the shear layer.

For a given $\theta_1$ and $l_1$, the separation region size scales with $l_2$, and hence the shear layer length $L_s$ increases as $\Lambda$ is increased. For small values of $\Lambda$, $L_s$ is sufficiently short for the shear layer to remain steady, as seen in the example shown in figure~\ref{fig:steadyFlow}. However, beyond a certain critical value of $\Lambda$, the shear layer has a sufficiently long development length for instabilities to manifest, pushing the entire flow into the oscillations regime of unsteadiness. During oscillations, unsteady flow structures in the shear layer induce small-amplitude high-frequency undulations in the intermediate shock wave structure. The separation region also experiences small-amplitude expansions and contractions in size due to impingement of the unsteady shear layer on the aft-cone surface. For intermediate values of $\Lambda$ the separation point $S$ also exhibits periodic fore-and-aft motion along the cone surface (see \S3.1). The dynamics of all these unsteady motions are coupled, and collectively the motions are termed as oscillations. The rest of this paper describes these oscillations in detail. 

\subsection{Effect of cone slant length ratio $\Lambda$: free- and anchored-oscillations \label{sec:lambda}}
\begin{figure}
	\centering
	\includegraphics[width=\linewidth]{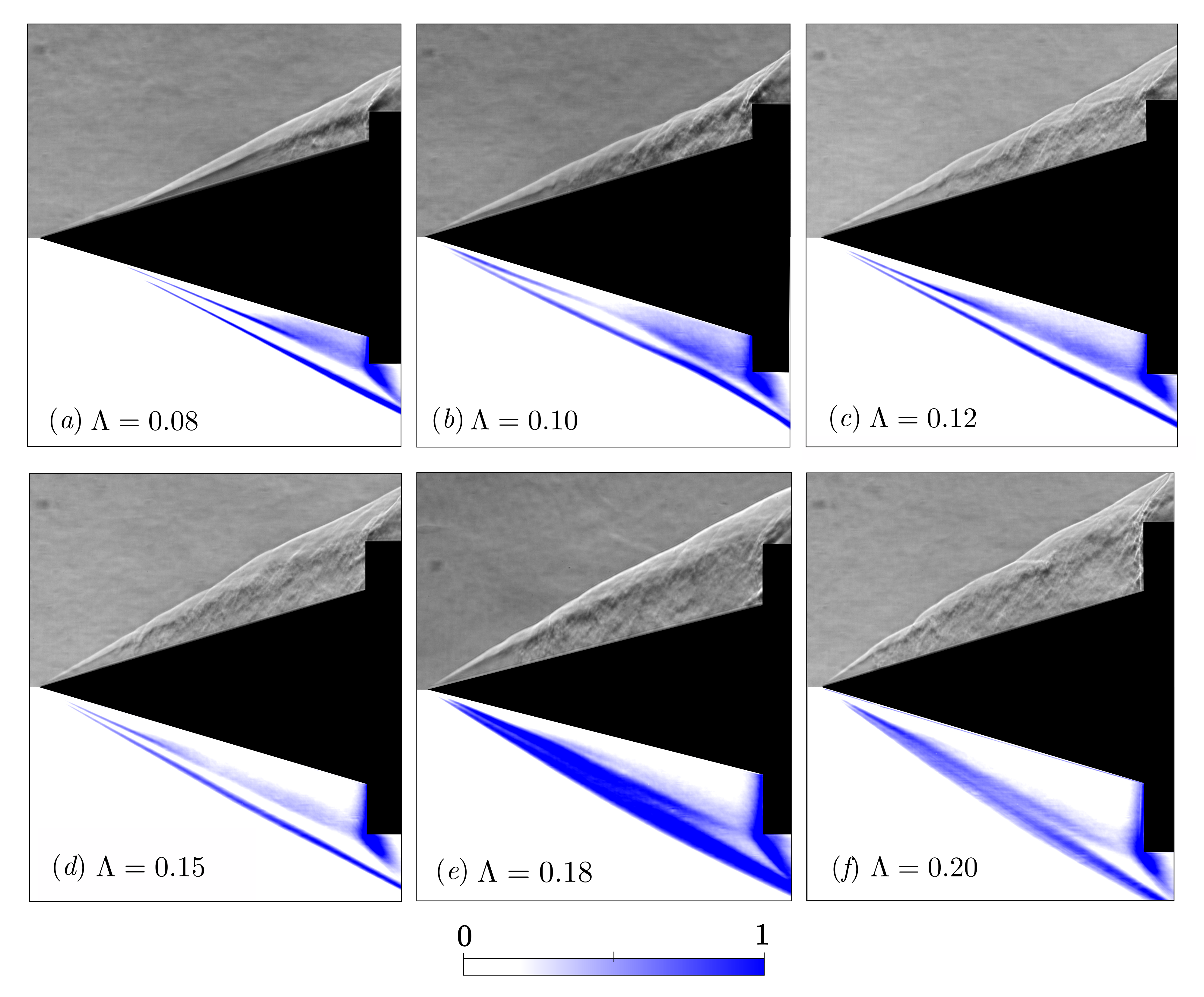}
	\caption{Instantaneous schlieren snapshots (top halves in gray scale) and normalized standard deviation of temporal fluctuations in schlieren image light intensity calculated from time-resolved data (bottom halves) for different values of $\Lambda$ at fixed $\theta_1 = 15^\circ$. White and blue colors represent low and high intensity of temporal fluctuations, respectively. Going from (\emph{a}) to (\emph{c}), it is seen that shear layer breakdown leads to increased thickness of the shear layer and higher levels of fluctuations. Lateral movement of the shear layer in anchored-oscillations cases (\emph{e} and \emph{f}) leads to a merger of the two regions that show large fluctuations, \emph{i.e.}, the areas around the separation shock wave and the shear layer.}
	\label{fig:15deg_SD}
\end{figure}

\begin{figure}
	\centering
	\includegraphics[width=\linewidth]{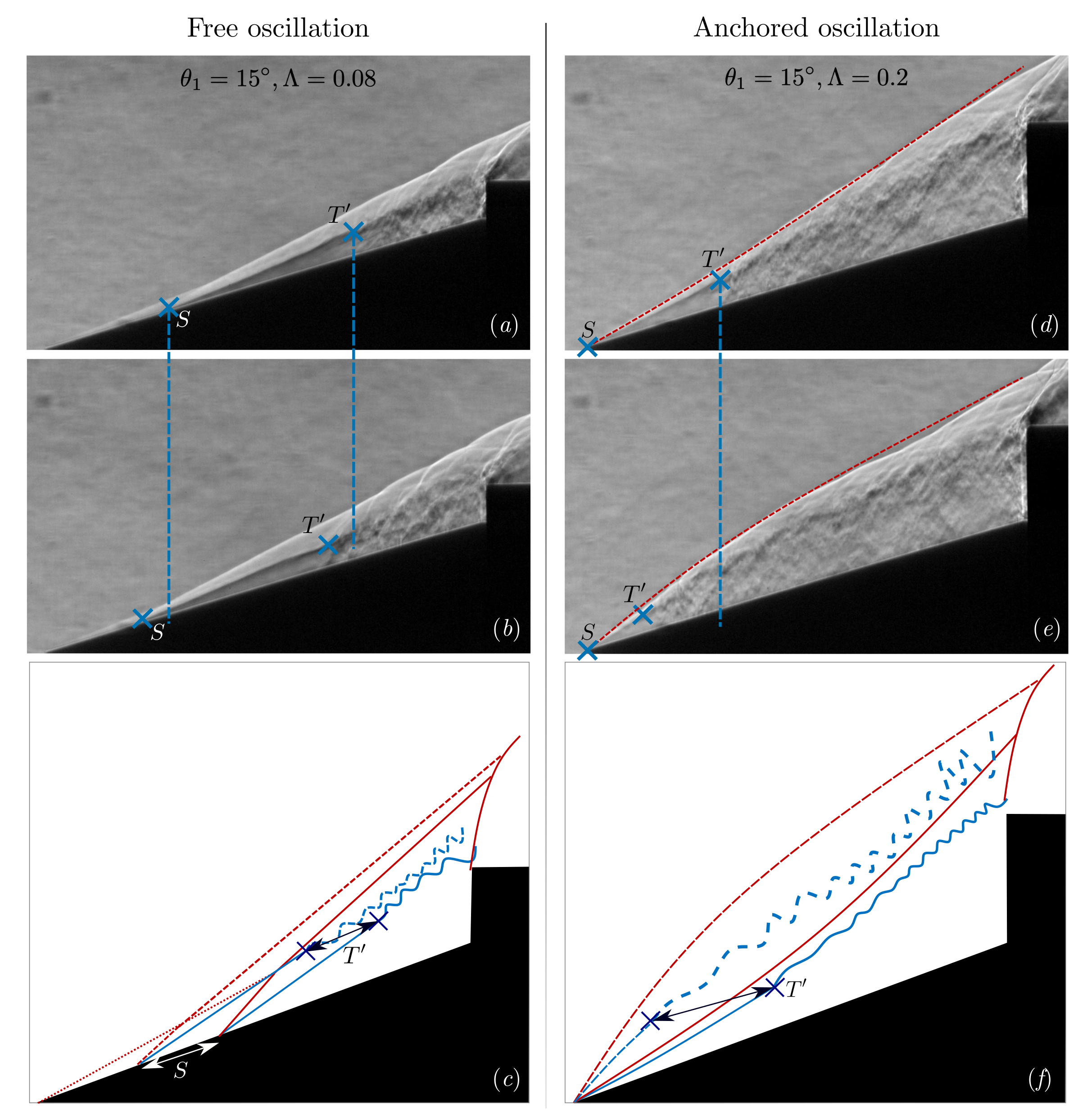}
	\caption{A comparison of flow features between free- and anchored-oscillations. The schlieren snapshots (\emph{a}) and (\emph{b}) are from a single cycle of free-oscillation, separated by approximately 180$^\circ$ in phase. Similarly, the schlieren snapshots (\emph{d}) and (\emph{e}) are from a single cycle of anchored-oscillation, separated by approximately 180$^\circ$ in phase. $T'$ marks the location where the shear layer breaks down due to rapid growth of instabilities. The schematics (\emph{c}) and (\emph{f}) illustrate the motion of the separation shock wave and the shear layer in free- and anchored-oscillations, respectively. The video provided in supplementary material also help in making a comparison between free- and anchored-oscillations.}
	\label{fig:DC_AnchoredOscillations}
\end{figure}

The effect of cone slant length ratio on the nature of oscillations is discussed here. As a representative example, figure~\ref{fig:15deg_SD} shows schlieren data for six different values of $\Lambda$ in the range 0.08 and 0.2 with fixed $\theta_1 = 15^\circ$. The influence of $\Lambda$ on flow behavior at other values of $\theta_1$ are qualitatively similar. The standard deviation maps of schlieren light intensity shown in the figure are generated from time-resolved data, and serve as a direct indicator of local flow unsteadiness levels. It is noted that schlieren light intensity is a linear measure of the local density gradient \citep{LR1957}. The growth in size of the separated flow region with increasing $\Lambda$ can be inferred from a visual comparison of the schlieren snapshots in the figure. A comparison of figures~\ref{fig:steadyFlow} and \ref{fig:15deg_SD}(\emph{a}) shows that as $\Lambda$ is increased from 0.06 to 0.08, there is a significant increase in the size of the separation region, and the downstream portion of the shear layer clearly becomes unsteady. From $\Lambda$ = 0.08 to 0.12, figures~\ref{fig:15deg_SD} (\emph{a}-\emph{c}), the size of the separation bubble increases in the lateral direction. An interesting aspect of flow oscillations for $\Lambda$ = 0.08, 0.10, 0.12 is the periodic fore-and-aft motion of the separation point $S$ along the fore-cone surface, and the accompanying motion of the separation shock wave. This aspect is highlighted in the left column of figure~\ref{fig:DC_AnchoredOscillations} where two snapshots from a single oscillation period with an approximate phase difference of $180^\circ$ are shown for $\Lambda = 0.08$. From a comparison of figures~\ref{fig:DC_AnchoredOscillations} (\emph{a}) and (\emph{b}) it is readily seen that the separation point $S$ translates along the cone surface, and the shear layer transition/breakdown point $T'$ also translates along with $S$ by approximately the same distance. The point $T'$ is visually identified in the schlieren images by the clear breakdown in the laminar shear layer structure. This type of flow oscillation, characterized by the periodic motion of the separation point, is labeled here as ``free-oscillation''.

As the size of the separation region grows with increasing $\Lambda$, the separation point ultimately reaches the cone nose for a sufficiently large $\Lambda$, and remains anchored in that location even as $\Lambda$ is further increased. In this regime of flow oscillations, where the separation point is anchored at the nose, the separation shock wave exhibits clear motion in the transverse direction due to periodic volume expansions (bulging) and contractions of the separation region. These motions, which are distinct from those observed in free-oscillations, can be understood in terms of the motion constraint on the separation point driving the separation region to expand in the transverse direction. This type of flow oscillation, characterized by the separation point remaining fixed at the cone nose, is labeled here as ``anchored-oscillation''. The right column of figure~\ref{fig:DC_AnchoredOscillations} shows a representative example of the same, where two snapshots from a single oscillation period with an approximate phase difference of $180^\circ$ are shown for $\Lambda = 0.2$. From a comparison of figures~\ref{fig:DC_AnchoredOscillations} (\emph{d}) and (\emph{e}) it is readily seen that the separation point $S$ remains anchored at cone nose, but interestingly, $T'$ shows a large excursion. This aspect of anchored-oscillations is discussed next.

In addition to the apparent geometrical differences in motions between free- and anchored-oscillations, an important distinction is seen in the instantaneous length of the shear layer between points $S$ and $T'$. It is noted that this length is simply the development length required by the shear layer to transition from a steady, laminar, state to an unsteady state under the given flow conditions. In free-oscillations, as $S$ translates along the cone surface, $T'$ follows the translations of $S$ such that the shear layer length between points $S$ and $T'$ remains approximately constant. This however is not the case in anchored-oscillations, as evidenced by the example shown in figures~\ref{fig:DC_AnchoredOscillations} (\emph{d}) and (\emph{e}). The large excursions of $T'$, which are periodic, can be explained using the stability properties of a compressible shear layer. For compressible shear layers \citep{brown1974density,papamoschou1988compressible}, it is known that the growth rate of instabilities in the shear layer region is dependent on Mach numbers on either sides of the layer ($M_o$ and $M_i$), Reynolds number ($Re = \rho_ou_oL/\mu_o$), density ratio ($\rho_o/\rho_i$), and the length of the shear layer to momentum thickness of the shear layer at the separation point ($L/\theta$). That is,
\begin{equation}
	\textup{Growth rate} = f(M_o,M_i, Re, \rho_o/\rho_i, L/\theta),
\end{equation}
where, $f$ denotes the functional dependence, and subscripts $o$ and $i$ refer to the locations above and below the shear layer, respectively (as shown in figure~\ref{fig:shearLayerScheme}). Since in the present case the Mach number on the low-speed side ($M_i$) is very small due to recirculating subsonic flow inside the separation bubble, the flow properties on the low-speed side can approximately be considered to be equal to the average local stagnation conditions. With that we have 
\begin{equation}
	M_i \rightarrow 0, \,\, \rho_o/\rho_i = g(M_o, \gamma), \,\textup{ and }\, Re = Re(M_o, P_i, T_i, L, \gamma), 
\end{equation}
where $g$ is obtained from basic isentropic flow relations. These simplifications lead to the conclusion that for a fixed stagnation flow conditions $P_i$, $T_i$ and $\gamma$ (which are constant in the wind tunnel experiments), the growth rate only depends on the local Mach number ($M_o$) on the high-speed side of the shear layer and length to momentum thickness ratio of the shear layer ($L/\theta$). 

Now, for free-oscillation cases where the separation point freely moves along the cone surface, it is observed that the separation shock wave angle and the shear-layer angle $\theta_s$ remain nominally constant throughout the oscillation cycle. This leads to a constant Mach number on the high-speed side of the shear layer, and hence a constant shear layer transition length as per the above arguments. The motions of the separated flow system in a free-oscillation cycle are schematically illustrated in figure~\ref{fig:DC_AnchoredOscillations}(c). In case of anchored oscillations, the shear layer transition location $T'$ experiences periodic motion even though the separation point $S$ remains anchored. The motion of $T'$ can be explained by the fact that the Mach number on the high-speed side of the shear layer is varying in a periodic manner. This variation is brought about by the periodic change in the local separation shock wave angle at the cone nose due to the bi-convex arching motion exhibited by the separation shock wave during the anchored-oscillation cycle. This bi-convex arching can be seen clearly in figures~\ref{fig:DC_AnchoredOscillations} (\emph{d} and \emph{e}), and the same is schematically illustrated in figure~\ref{fig:DC_AnchoredOscillations} (\emph{f}). From basic shock wave relations we know that higher the shock wave angle, lower the downstream Mach number. Hence we expect the Mach number above the shear layer ($M_o$) to be the smallest at the time instant when the separation shock wave has the maximum outward bulge (largest shock wave angle locally at the nose), and the Mach number to be the largest at the time instant when the separation shock wave has the maximum inward depression (smallest shock wave angle locally at the nose). In general, the stability of compressible shear layers increases with Mach number, \emph{i.e.}, at a higher Mach number, instabilities in the shear layer manifest at higher values of local flow Reynolds number, needing a longer development length for shear layer breakdown \citep{papamoschou1988compressible}. Hence we expect the shear layer to have a longer development length for transition at the time instant when the separation shock wave is depressed inward as compared to the time instant when the shock wave is bulging outward. The experimental observations in figures~\ref{fig:DC_AnchoredOscillations} (\emph{d} and \emph{e}) are very much consistent with this expectation.

Among the oscillations cases shown in figure~\ref{fig:15deg_SD}, $\Lambda = 0.15$ falls at the cusp, between free- and anchored-oscillation states of unsteadiness. For $\Lambda = 0.15$ the separation point $S$ is seen to be anchored to the cone nose for a part of the oscillation cycle period, and for the rest of the period $S$ exhibits a small excursion from the nose. We refer to these boundary cases as mixed-oscillations, since they contain features of both free- and anchored-oscillation states. For $\Lambda = 0.18 \textup{ and } 0.20$ however, the flow completely transitions to a state where the separation point is fully anchored to the cone nose for entire oscillation cycle and shows shear layer and shock wave movement as discussed above. In summary, for a fixed $\theta_1$ and starting at a small value of $\Lambda$ where the flow is steady with a small separation bubble, an increase in $\Lambda$ transitions the flow to a state of free-oscillations, and subsequently to a state of anchored-oscillations. On further increasing $\Lambda$, the flow transitions to a pulsation state, which is fundamentally very different from oscillations, and is not of present concern.

\subsection{Effect of fore-cone angle \textup{(}$\theta_1$\textup{)} \label{sec:theta1}}
\begin{figure}
	\centering
	\includegraphics[width=0.85\linewidth]{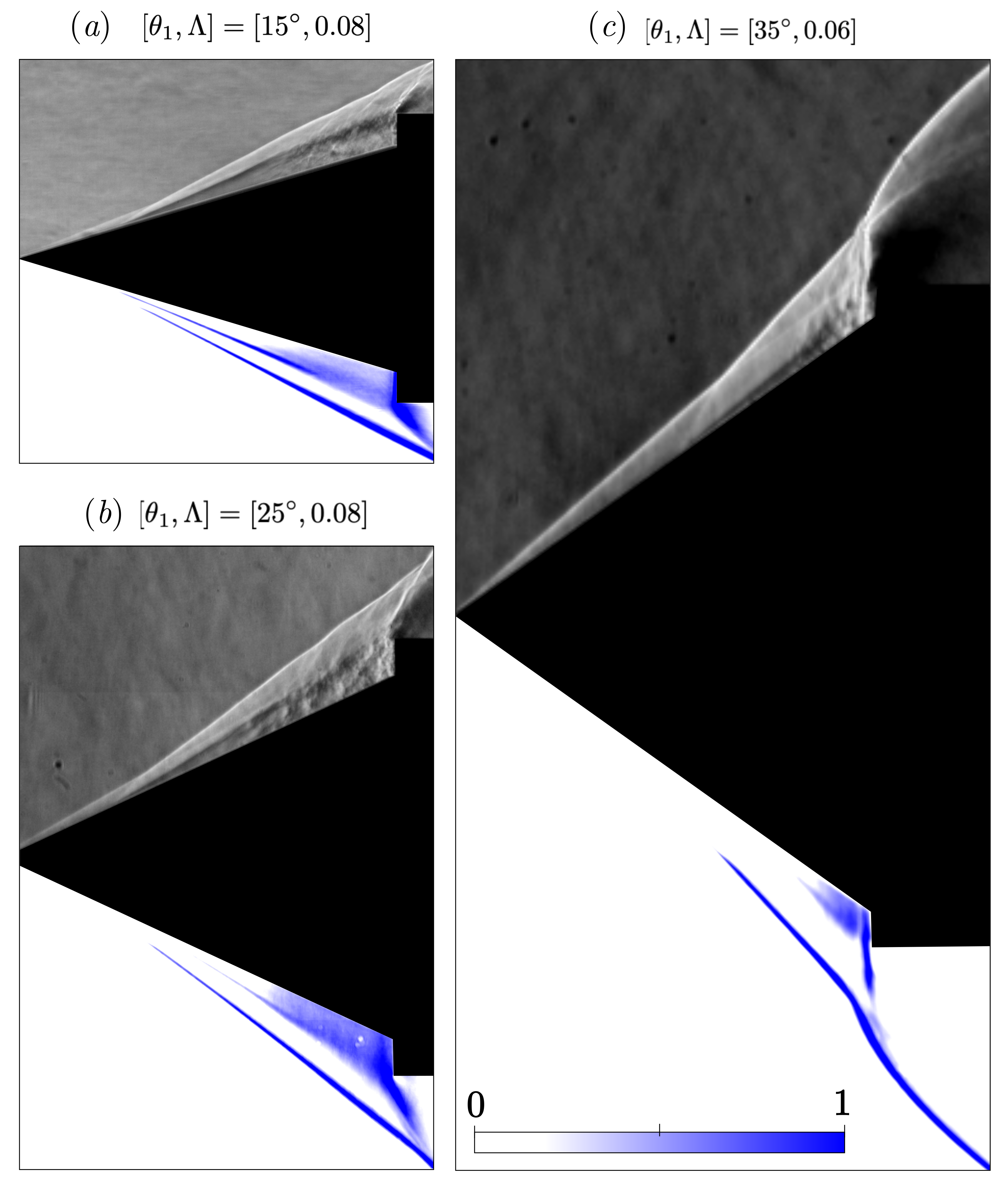}
	\caption{Instantaneous schlieren snapshots (top halves of sub-figures) and standard deviation maps of schlieren light intensity (bottom halves of sub-figures).}
	\label{fig:25deg_SD}
\end{figure}

To illustrate the effects of fore-cone angle ($\theta_1$) on the flow, cases with $\theta_1 = 15^\circ$, $25^\circ$, $35^\circ$ are briefly discussed here as representative examples. Figure~\ref{fig:25deg_SD} shows schlieren data for the three different angles. A visual comparison across the sub-figures in figure~\ref{fig:25deg_SD} shows that an increase in $\theta_1$, with approximately the same $\Lambda$, brings about a reduction in the size of the separation region. It is noted that the flow turn angle ($90^\circ$-$\theta_1$) at the vertex between the two cones decreases with increasing $\theta_1$, and it is expected that a lower turn angle results in a smaller separation region since pressure gradients generated near the wall by the turning flow will be less adverse.

Another effect of increasing $\theta_1$ is the increase in the conical shock wave angle, and the resulting decrease in the Mach number downstream of the conical shock wave and above the shear layer ($M_o$). As discussed above, a decrease in $M_o$ leads to shear layer breakdown at a relatively upstream location, \emph{i.e.}, a shorter development length is needed for shear layer transition. This effect is seen when figures~\ref{fig:25deg_SD} (\emph{a}) and (\emph{b}) are compared, where $\Lambda = 0.08$ in both cases, and also in comparison with figure~\ref{fig:25deg_SD} (\emph{c}) where $\Lambda$ is nearly the same.

\subsection{Strouhal number scaling}
\begin{figure}
	\centering
	\includegraphics[width=0.7\linewidth]{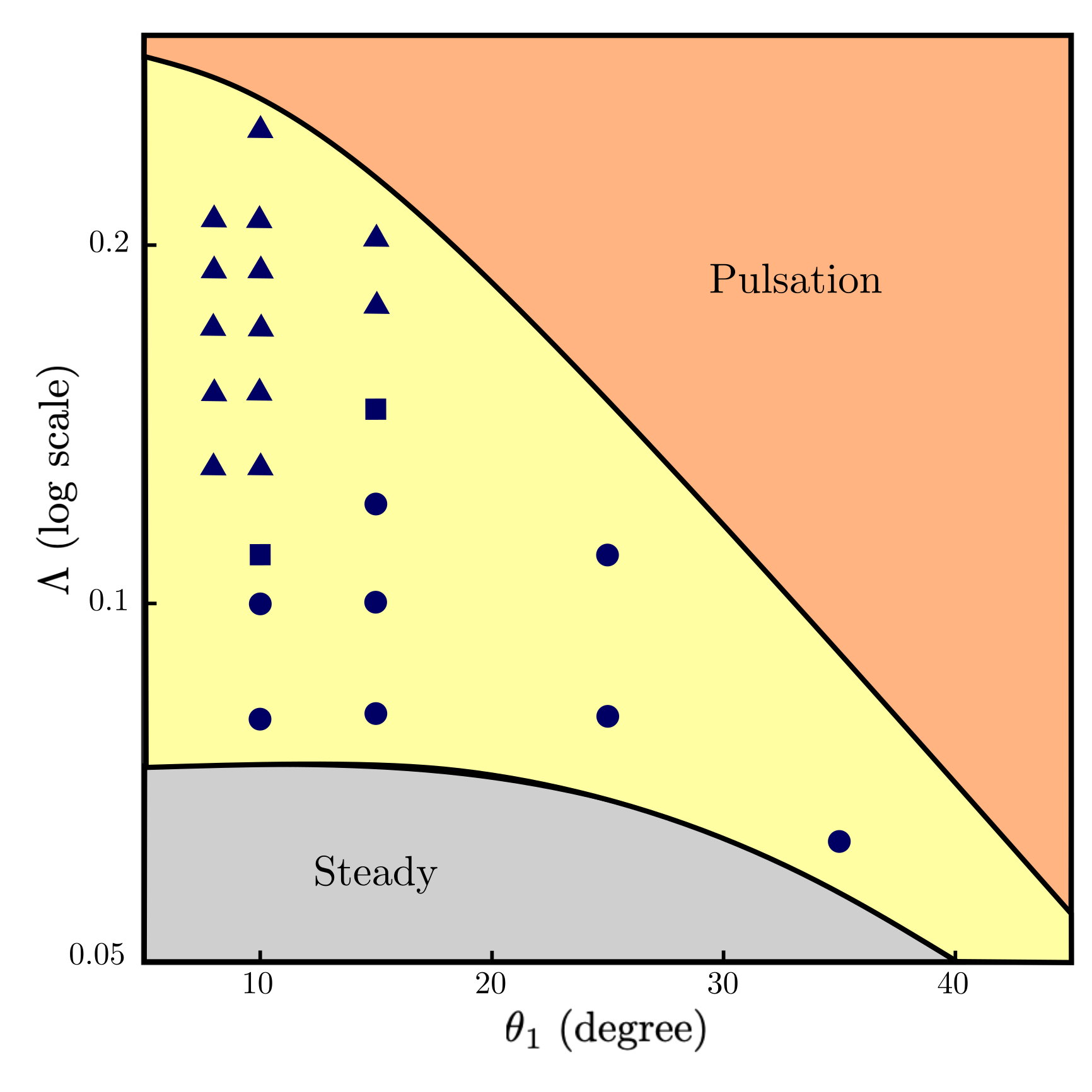}
	\caption{The $\theta_1$-$\Lambda$ parameter space for the double cone flow with $\theta_2 = 90^\circ$. The same set of experimental data markers as figure~\ref{fig:mounted_model} (\emph{b}) are shown here, but the markers are now classified as per oscillation types discussed in \S3.1: free-oscillations ($\bullet$); anchored-oscillations ($\blacktriangleup$); mixed-oscillations that show features of both free- and anchored-oscillations ($\sqbullet$).}
	\label{fig:AllExperiments}
\end{figure}

The $\theta_1$-$\Lambda$ parameter space shown in figure~\ref{fig:mounted_model} (\emph{b}) is shown again in figure~\ref{fig:AllExperiments}, but here the experimental data markers are classified as per the oscillation types discussed above in \S3.1, \emph{i.e.}, free-oscillations, anchored-oscillations, and mixed-oscillations that show features of both free- and anchored-oscillations. From the figure, it is noted that only free-oscillations were observed for $\theta_1 = 25^\circ$ and $35^\circ$ at the $\Lambda$ values chosen for this study, whereas for $\theta_1 \leq 15^\circ$ both sub-types of oscillations, free- and anchored-oscillations, were observed. 

\begin{figure}
	\centering
	\includegraphics[width=0.8\linewidth]{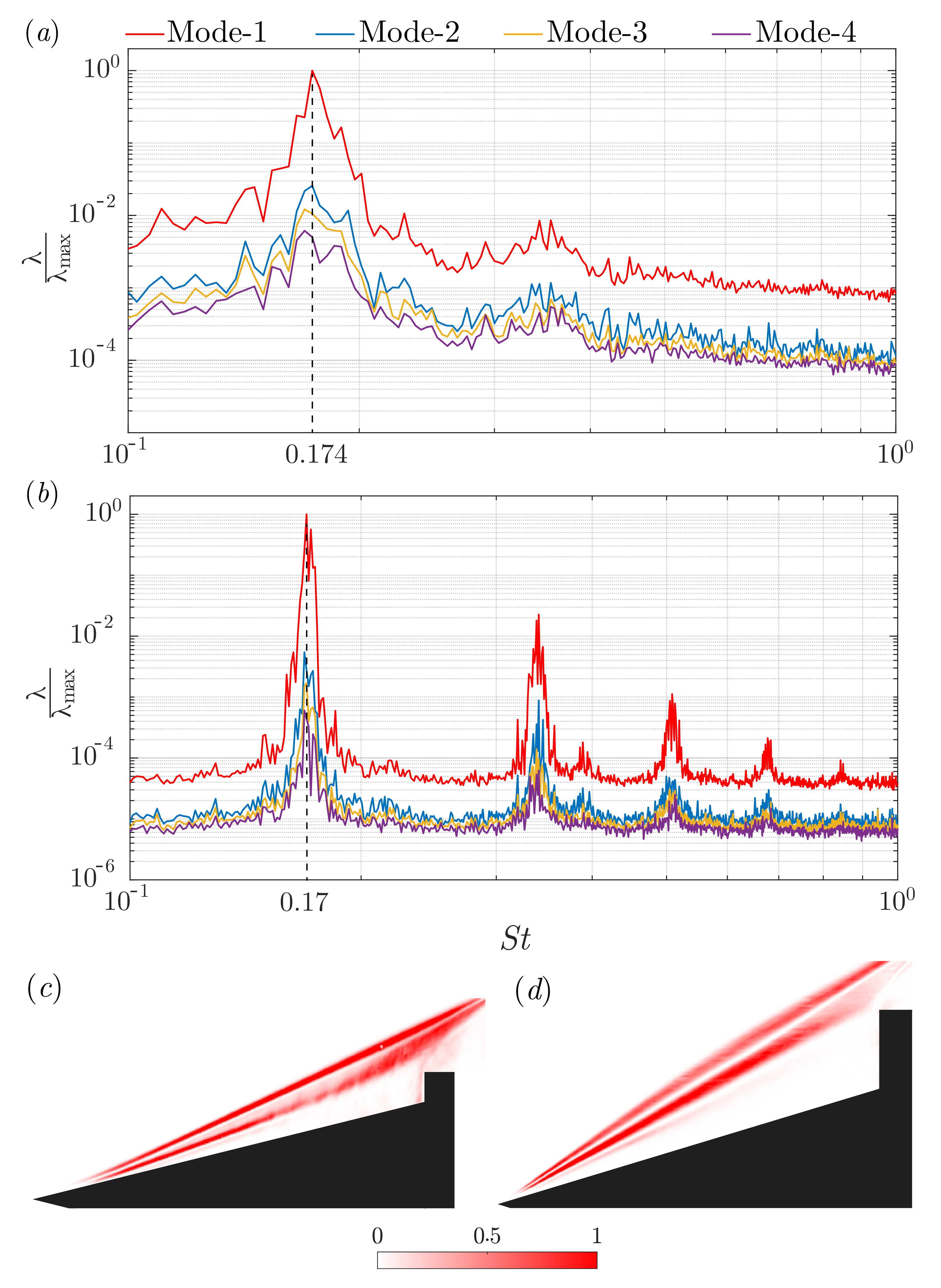}
	\caption{SPOD modal energy spectra normalized by maximum energy of the leading mode and the corresponding mode shape of the leading mode (bottom row). (\emph{a}, \emph{c}) Free-oscillation for $\theta_1=15^\circ$ and $\Lambda=0.08$. (\emph{b}, \emph{d}) Anchored-oscillation for $\theta_1=15^\circ$ and $\Lambda=0.20$.}
	\label{fig:SPOD}
\end{figure}

Qualitative observations from all the schlieren data obtained for flow oscillations in the present work suggest that the unsteady motions of the shear layer, the separation point, and the separation and intermediate shock waves are coupled, and contain a single dominant time scale. To extract such a global time scale from the time-resolved data, we apply the technique of spectral proper orthogonal decomposition (SPOD) \citep{towne_schmidt_colonius_2018}. SPOD, which is a frequency-domain variant of the conventional proper orthogonal decomposition (POD), is an effective tool for extracting key temporal scales from time-resolved schlieren data where the flow unsteadiness is statistically stationary \citep{TD2022cylinder}. To briefly summarize SPOD implementation with schlieren data, consider an ensemble of realizations of schlieren intensity field $q(\bold{x},t) = \{q_j(\bold{x},t),\, j=1,2,\dots,n\}$, where $n$, $\bold{x}$, and $t$ are the total number of realizations, spatial coordinate vector, and time, respectively. The spatio-temporal field $q(\bold{x},t)$ is first transformed to power spectral density (PSD) field $\hat{q}(\bold{x},f) = \{\hat{q}_j(\bold{x},f),\, j=1,2,\dots,n\}$ in the frequency ($f$) domain, and subsequently, $\hat{q}(\bold{x},f)$ is decomposed into orthogonal modes at individual frequencies through POD:
\begin{equation}
	\hat{Q}(\bold{x},f) = \sum_{j=1}^n a_{j}(f)\psi_{j}(\bold{x},f).
	\label{eq:SPOD_decomp}
\end{equation}
This essentially means that SPOD decomposes the ensemble of realisations $\hat{q}(\bold{x},f)$ into orthogonal modes $\psi_{j}(\bold{x},f)$  at different frequencies, and $\hat{Q}(\bold{x},f)$ represents the optimal reconstruction of the flow data set that contains the maximum energy (variance) of the ensemble. This optimization problem leads to an eigen value decomposition problem, and the eigen values corresponding to these modes at different frequencies are represented by $\lambda_j(f) = a_{j}^2(f)$.

\begin{figure}
	\centering
	\includegraphics[width=\linewidth]{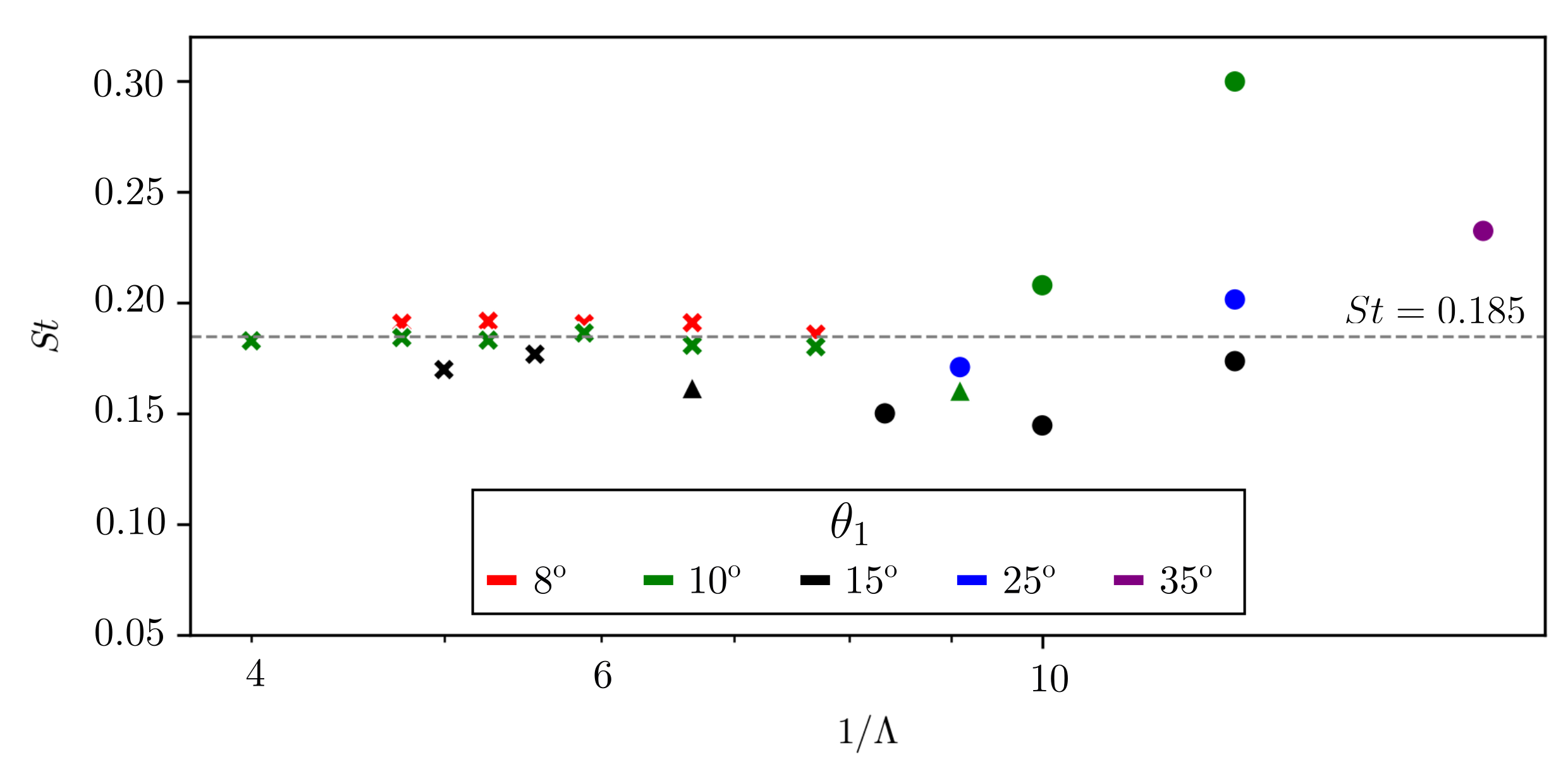}
	\caption{Strouhal number ($St = f l_1/U_\infty$) variation with $\theta_1$ and $\Lambda$. Marker shapes represent oscillations sub-types: free-oscillation ($\bullet$); anchored-oscillation ($\bold{\times}$); mixed-oscillation ($\blacktriangleup$). Marker color indicates $\theta_1$ as per the inset figure legend.}
	\label{fig:Strouhalno}
\end{figure}
\begin{figure}
	\centering
	\includegraphics[width=\linewidth]{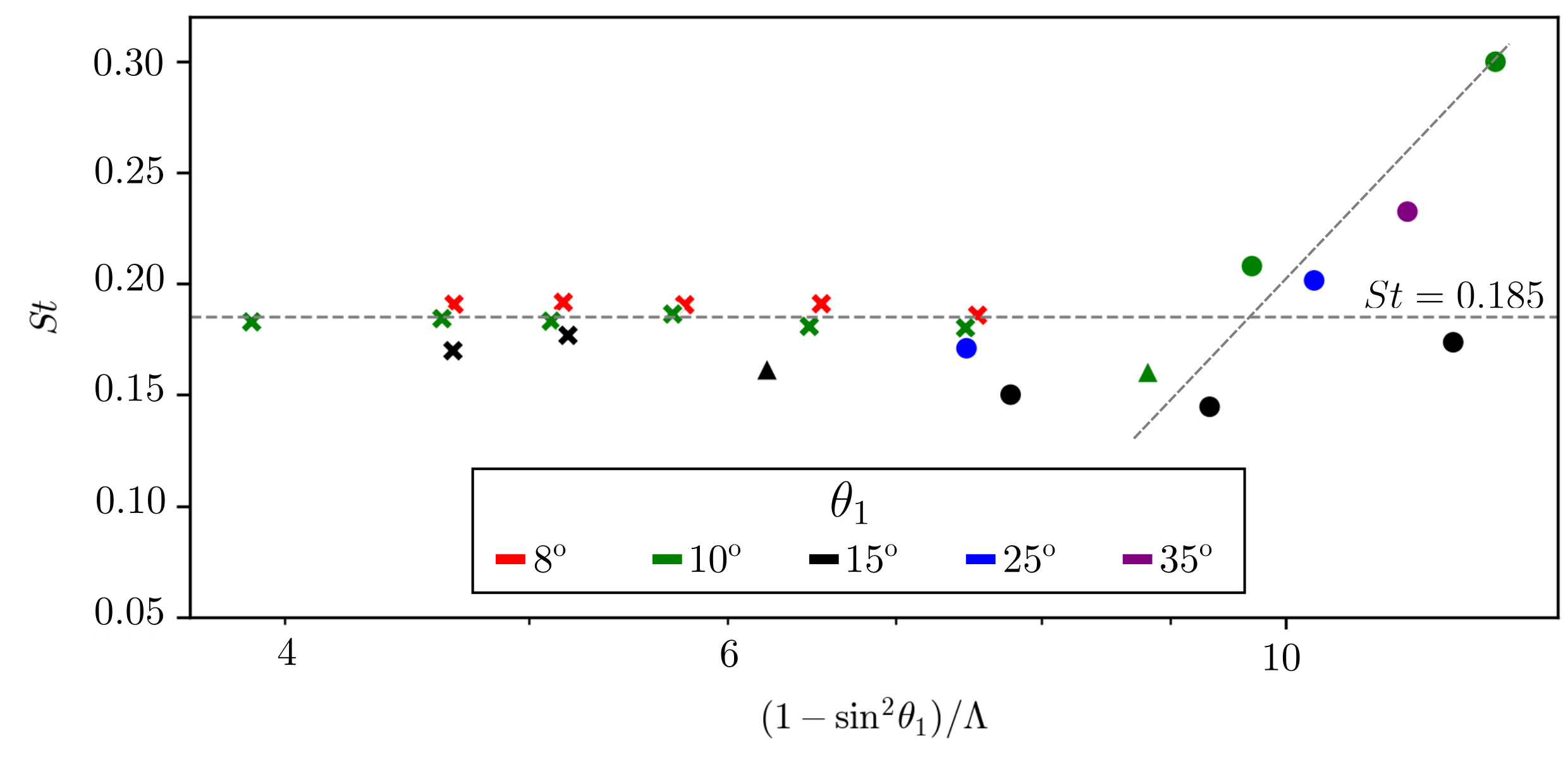}
	\caption{Strouhal number ($St = f l_1/U_\infty$) variation with $\theta_1$ and $\Lambda$. The horizontal axis is scaled with the parameter $(1-\textup{sin}^2\theta_1)/\Lambda$ which combines $\theta_1$ and $\Lambda$. Figure legend same as that of figure~\ref{fig:Strouhalno}.}
	\label{fig:Strouhalno1}
\end{figure}

For the present data set, PSD of the schlieren snapshot matrix was calculated to obtain convergent estimates of spectral density as per the recommendation of \citet{towne_schmidt_colonius_2018}. The SPOD modes are ranked based on the corresponding $\lambda_j$ values, and then the leading mode can be understood to represent the dominant coherent structure in the flow. Figure~\ref{fig:SPOD} shows SPOD results for two specific oscillation cases discussed in \S3.1 (see figure~\ref{fig:DC_AnchoredOscillations}). In both the cases the leading mode is seen to be much more energetic in comparison with the other modes, and hence it provides a good representation of the coherent spatial structure present in the flow. A peak in the spectra can be clearly identified in both cases, which indicates the presence of a dominant global time scale that characterizes flow unsteadiness, \emph{i.e.} a characteristic frequency for oscillations. A non-dimensional frequency $St$ (Strouhal number) is defined here as
\begin{equation}
	St=\frac{f\,l_1}{U_\infty}\, ,
        \label{eq:strouhal}
\end{equation}
where $U_\infty$ is the free-stream velocity. From the spectra plots, the Strouhal numbers for these free- and anchored-oscillations cases are inferred to be 0.174 and 0.170, respectively. The SPOD analysis outlined here was performed for all the data markers shown in figure~\ref{fig:AllExperiments}. The Strouhal number thus obtained for different combinations of $\theta_1$ and $\Lambda$ are summarized in figure~\ref{fig:Strouhalno}.

For anchored-oscillations, the Strouhal number seems to be mostly invariant with respect to both $\theta_1$ and $\Lambda$, and shows an average value of $St = 0.185$. Whereas for free-oscillations the Strouhal number exhibits a clear dependency on both $\theta_1$ and $\Lambda$. The possible reason for the Strouhal number behavior observed in figure~\ref{fig:Strouhalno} is discussed in the following section. Figure~\ref{fig:Strouhalno1} shows the Strouhal number data scaled by the parameter $(1-\textup{sin}^2\theta_1)/\Lambda$ which combines $\theta_1$ and $\Lambda$. With this scaling the data for free-oscillations appears to scale linearly with $(1-\textup{sin}^2\theta_1)/\Lambda$. In this context it is noted that Mach number $M_o$ decreases with increase in either $\Lambda$ or $\theta_1$ (as seen previously in subsections \ref{sec:lambda} and \ref{sec:theta1}). Hence $M_o$ monotonically increases with the parameter $(1-\textup{sin}^2\theta_1)/\Lambda$. The scaling observed here suggests that $M_o$ could be an important parameter in determining $St$, and this aspect is naturally incorporated in the aeroacoustic model developed in the following section. 

\section{An aeroacoustic model for oscillations \label{sec:mechanism}}
\begin{figure}
	\centering
    \includegraphics[width=0.8\linewidth]{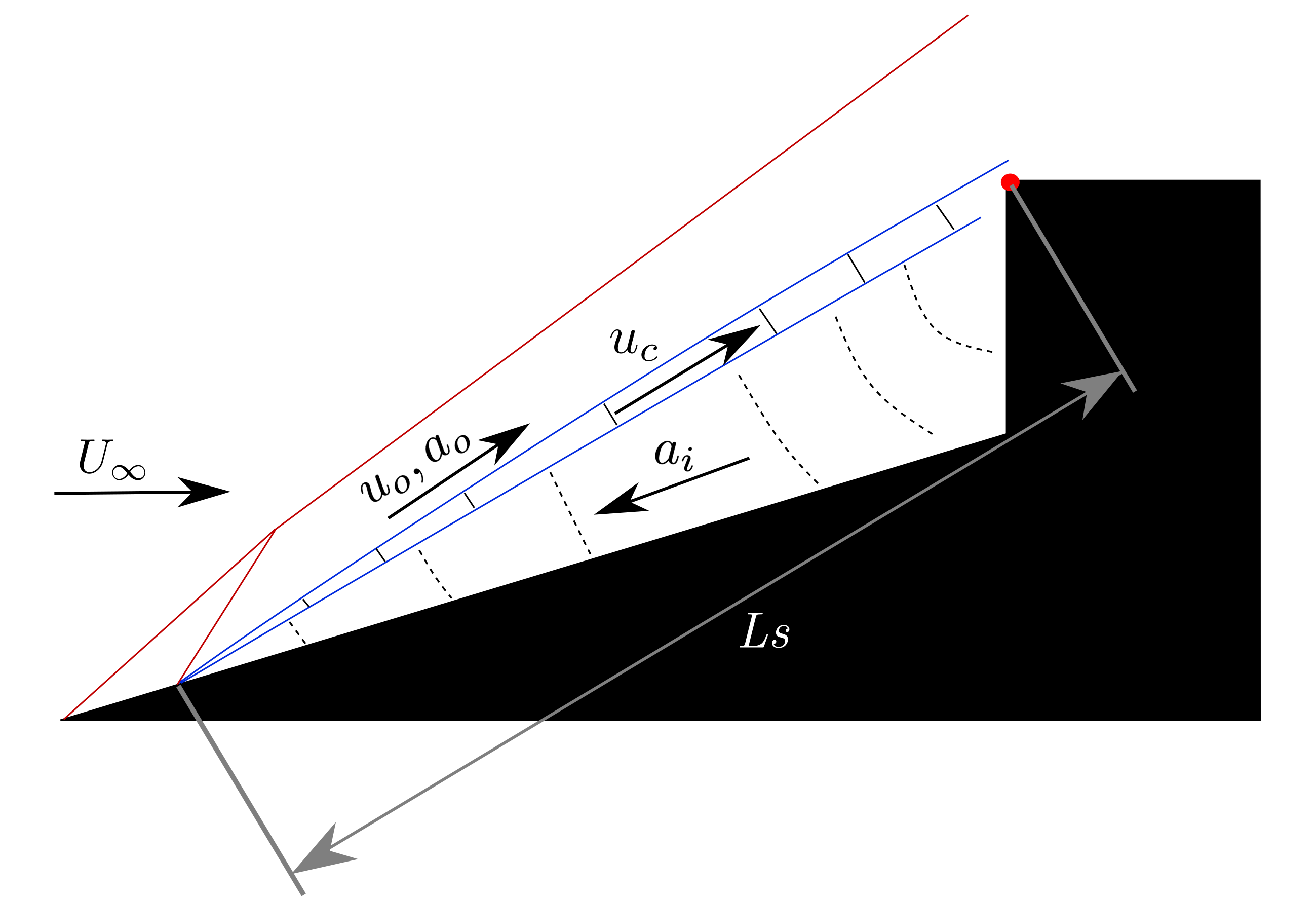}
	\caption{A schematic representation of the proposed aeroacoustic feedback mechanism in the double cone model.}
	\label{fig:DCrossiter}
\end{figure}
\begin{figure}
	\centering
	\includegraphics[width=0.75\linewidth]{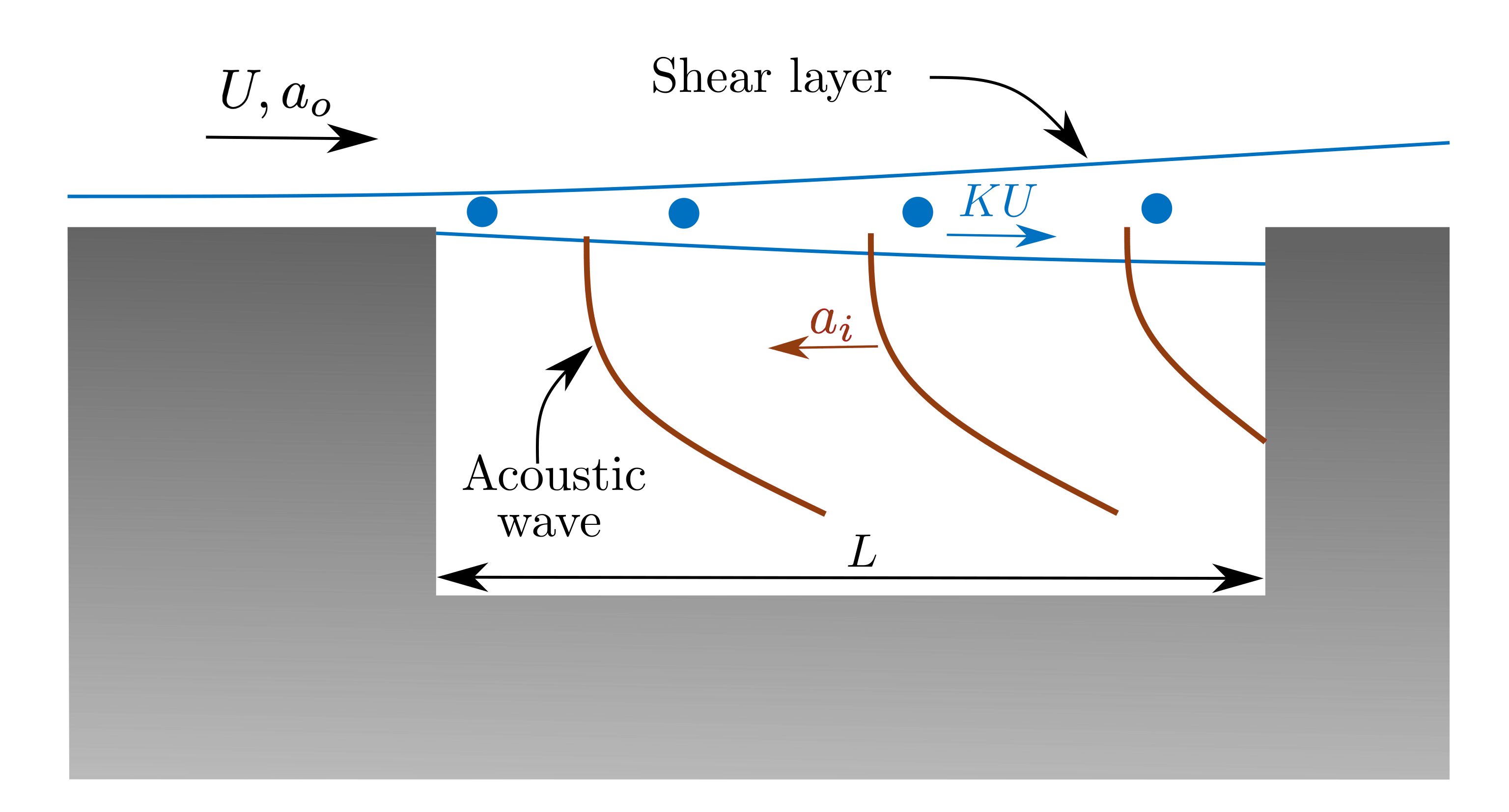}
	\caption{A schematic representation of the aeroacoustic feedback cycle in an open cavity flow. The disturbances in the shear layer propagate at a fraction of the free-stream speed ($KU$). The shear layer impingement at the trailing edge of the cavity generates acoustic waves that propagate upstream in the separated flow region to influence the shear layer at the leading edge of the cavity. }
	\label{fig:rossiter}
\end{figure}

The mean flow structure of oscillations over a double cone is shown in a simple schematic manner in figure~\ref{fig:DCrossiter}. As discussed in the previous sections, the flow configuration comprises of a compressible shear layer sitting over a separation bubble, and impinging at the aft-cone shoulder. This entire configuration exhibits self-sustained coherent flow oscillations in time. It is noted that these flow features have broad similarities to compressible flow over an open cavity, illustrated schematically in figure~\ref{fig:rossiter}. Both the double cone and open cavity configurations comprise of a large separated flow region with a compressible free-shear layer, and exhibit low-frequency flow unsteadiness. The phenomena of flow oscillations in compressible open cavity flows has been extensively studied in the past \citep{rossiter1964wind, heller1971flow, rockwell1978self, tam1978tones, rowley2002self, larcheveque2003large, kegerise2004mode, zhuang2006supersonic, bres2008three}, and it is largely explained through an aeroacoustic feedback mechanism that was first described in detail by \cite{rossiter1964wind}. This feedback mechanism is composed of two phases in a flow cycle (refer to figure~\ref{fig:rossiter}): (1) the downstream propagation and growth of disturbances in the compressible shear layer; (2) the upstream propagation of acoustic disturbances in the subsonic separated flow region. In this feedback cycle, disturbances at the leading edge of the cavity are excited by the upstream propagating acoustic disturbances, and in turn, the acoustic disturbances at the trailing edge of the cavity are generated due to the impingement of the shear layer disturbances on the cavity trailing edge. Based on this aeroacoustic mechanism, \cite{rossiter1964wind} proposed the following equation to determine the resonant frequencies of cavity flow oscillations:
\begin{equation}
	f = \frac{1}{L}\left(\frac{m-\kappa}{\frac{1}{KU}+\frac{1}{a_i}}\right).
	\label{eq:RossiterFormula}
\end{equation}
Equation \eqref{eq:RossiterFormula} essentially says that in a cavity of length $L$, the time period of oscillation ($1/f$) is set by the feedback loop which comprises of downstream propagating disturbances at a speed proportional to free-stream speed ($U$) and upstream propagating disturbances at acoustic wave speed ($a_i$). Here, $m$ represents the mode (or overtone) of the natural resonance frequency and $\kappa$ represents the phase lag between the shear layer disturbance wave and acoustic wave near the trailing edge of the cavity. The proportionality constant ($K$) has been determined in the literature based on experimental data. Drawing inspiration from the open cavity flow, and based on qualitative observations from our time-resolved schlieren data, we hypothesize that an aeroacoustic feedback loop could give rise to oscillations in the double cone flow. Based on this hypothesis, a model to predict the oscillation frequency is constructed in the following manner.

\begin{figure}
	\centering
	\includegraphics[width=\linewidth]{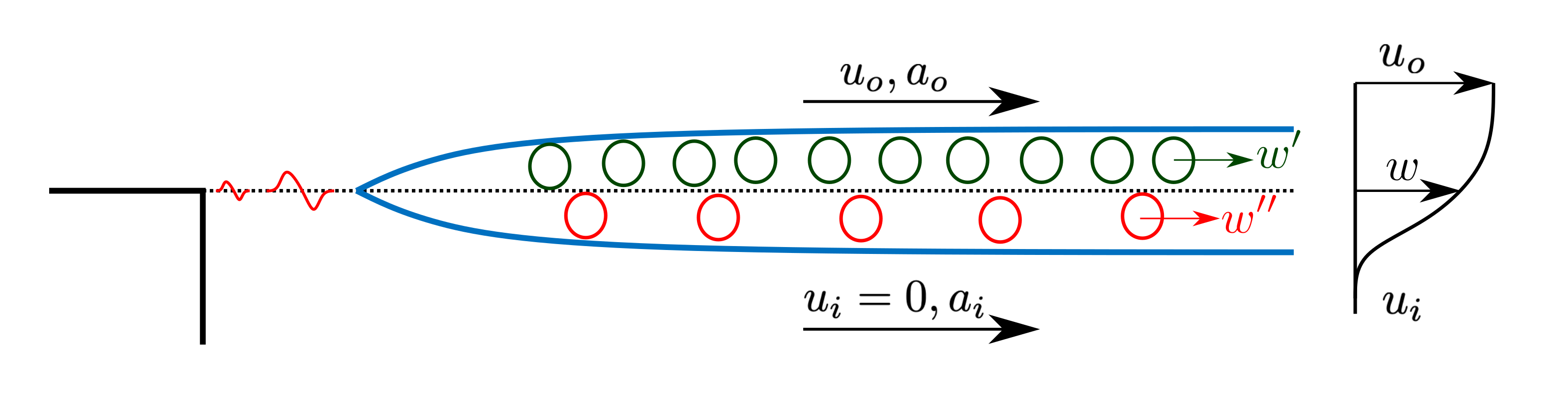}
	\caption{A schematic representation of the compressible shear layer (separated boundary layer) based on the vortex train model proposed by \cite{oertel2016mach}. The flow speeds on either sides of the shear layer are $u_{i}$ (= 0) and $u_o$, corresponding to the stagnant flow in the separation bubble and supersonic flow above the shear layer, respectively.}
	\label{fig:ortelsen}
\end{figure}

We consider disturbances in the compressible shear layer to propagate downstream at a speed $u_c$, which is dependent on the flow speed above the shear layer $u_o$, but not necessarily proportional to $u_o$ as assumed in many of the cavity flow studies at low supersonic Mach numbers. To obtain $u_c$, we rely on the significant amount of literature available on compressible shear layers, where the flow structure and properties such as growth rate of the shear layer thickness, convection speed of disturbances, growth rate of dominant waves and their structure, \textit{etc}. has been studied in detail \citep{bogdanoff1983compressibility, papamoschou1988compressible, elliott1990compressibility, hall1993experiments, murray2001characteristics, pantano2002study}. According to \cite{papamoschou1988compressible}, the disturbance speed in the shear layer $u_c$ can be related to $u_o$ as 
\begin{equation}
	u_c = \frac{u_o}{1+\frac{a_o}{a_i}},
	\label{eq:papamoschou}
\end{equation}
where the flow speed below the shear layer (in the separation region) is assumed to be zero. Here, $a_o$ and $a_i$ are the speeds of sound on the high- and low-speed sides of the shear layer, respectively. Extensive experimental studies by Ortel sen on supersonic jets show that disturbances in a compressible shear layer are primarily comprised of three types, each propagating downstream with a distinct characteristic speed \citep{oertel1979mach,oertel1980mach,oertel1983coherent}. Subsequently, the existence of these three types of disturbance waves was shown by \cite{tam1989three} through harmonic analysis of linearized governing equation of compressible inviscid flow. Some important features of the disturbance waves are briefly summarized here. The first type of disturbance is the Kelvin-Helmholtz (K-H) instability wave, and the second and third types of disturbances are termed as supersonic and subsonic instability waves, respectively. The supersonic instability wave is centered around the shear layer centerline and propagates downstream at a speed slower than K-H instability wave, which is centered in the high-speed side of the shear layer. The subsonic instability mode is centered in the low-speed side of the shear layer and can propagate upstream in the subsonic flow regions of the shear layer. In general, K-H and supersonic instability waves show weak dependence on the shear layer thickness, whereas the subsonic wave is sensitive to the shear layer thickness and exhibits neutral stability for the theoretical case of zero shear layer thickness. Based on the theoretical and experimental understanding of the disturbance waves present in a compressible shear layer \citep{tam1971directional,tam1989three,tam2009mach,sen2010new}, a vortex train model was proposed by \cite{oertel2016mach} to obtain the propagation speeds for the three types of disturbances. For a shear layer with flow speed $u_o$ on the high-speed side and stationary fluid on the low-speed side, the propagation speed can be expressed as: 
\begin{equation}
	\begin{split}
		&w = \frac{u_o}{1+r}, \qquad r = \frac{a_o}{a_i},\\
		&w' = \frac{u_o+rw}{1+r},\\
		&w'' = \frac{u_o-rw}{1+r}.
	\end{split}
	\label{eq:OrtelsenModel}
\end{equation}
In the above equation $w$, $w'$ and $w''$ represent propagation speeds of supersonic, K-H, and subsonic instability waves. Here, $a_o$ and $a_i$ are the speeds of sound on the high- and low-speed sides of the shear layer, respectively.  A more detailed derivation of expressions in equation \eqref{eq:OrtelsenModel} can be found in  \cite{oertel2016mach}. A schematic representation of the compressible shear layer based on the vortex train model of \cite{oertel2016mach} is shown in figure~\ref{fig:ortelsen}. It is noted here that the speed $u_c$ obtained from equation \eqref{eq:papamoschou} is same as the speed $w$ obtained from equation \eqref{eq:OrtelsenModel}. Thus, for a known flow speed $u_o$, and $a_o/a_i$, the downstream propagation speeds of the three types of disturbances in the compressible shear layer can be estimated. 

In the double cone flow, acoustic waves are generated when the unsteady shear layer impinges at the shoulder of the aft-cone model, and the acoustic waves can travel upstream inside the subsonic separated flow region. This is very similar to the aeroacoustic mechanism that is at play in the open cavity flow. Considering that the fluid inside the separation bubble is nearly stagnant, the acoustic wave propagation speed $a_i$ can be estimated from the average local temperature in the bubble. In supersonic cavity flow literature, the average local temperature is considered same as the stagnation temperature $T_0$ \citep{rockwell1978self}. This assumption however does not hold very well at high supersonic Mach numbers ($M_o = u_o/a_o$) due to high viscous dissipation in the shear layer. A better estimate for average local temperature in the separation bubble can be obtained by calculating the recovery temperature inside the separation bubble which accounts for the viscous losses in the shear layer \citep{anderson2016ebook}
\begin{equation}
	\begin{split}
		a_i &= \sqrt{\gamma R T_i}\,,\\
		T_i &= \frac{1 \,+\, \sqrt{Pr} \left(\frac{\gamma-1}{2}\right) M_o^2}{1\,+\,\left(\frac{\gamma-1}{2}\right)M_o^2\,}\,\, T_0\,,
	\end{split}
\label{eq:arEquation}
\end{equation}
where $T_0$ is the flow stagnation temperature in the free-stream, and $\gamma$ and $Pr$ are the ratio of specific heats and the Prandtl number, respectively, for air.

We can now use the ansatz of Rossiter's model to work out an expression for the Strouhal number of flow oscillations over a double cone. From equations~\ref{eq:strouhal} and \ref{eq:RossiterFormula}, we have
\begin{equation}
	\begin{split}
		St \,=\, \frac{f l_1}{U_\infty} &\,=\, \left(\frac{l_1}{U_\infty}\right)\left(\frac{1}{L_s}\right)\left(\frac{m-\kappa}{\frac{1}{u_c}+\frac{1}{a_i}}\right)\\
		& \,=\, \left(\frac{1}{M_\infty}\right)\left(\frac{a_o}{a_\infty}\right)\left(\frac{l_1}{L_s}\right)\left(\frac{m-\kappa}{\frac{1}{KM_o}+\frac{a_o}{a_i}}\right).
	\end{split}
\label{eq:dcrossiter}
\end{equation}
Here $L_s$ is the shear layer length measured from the separation point to the reattachment point. In the present model, the disturbance propagation speed $u_c$ can take one of three values ($w$, $w'$ or $w''$). Here the parameter $K$ is defined as the ratio $u_c/u_o$, which according to equation \eqref{eq:OrtelsenModel} is a function of $r = a_o/a_i$ (unlike open cavity flow models where it is taken to be a constant). The dominant frequency is obtained with the first mode of oscillation ($m = 1$), and the value of $\kappa$ is determined by the phase difference between the downstream propagating disturbances in the shear layer and the acoustic wave generated at the shoulder of the double cone model. In the present model, $\kappa = 0.25$ corresponding to a phase difference of $90^\circ$ is assumed based on the previous literature on open cavity flows \citep{rockwell1978self}. The ratios $a_o/a_i$, $a_o/a_\infty$ and $K$ are obtained using the following expressions:
\begin{equation}
\label{eq:parameters}
	\begin{split}
		\frac{a_o}{a_i} &\,=\, \frac{1}{\left[1 \,+\, \sqrt{Pr} \left(\frac{\gamma-1}{2}\right) M_o^2\right]^\frac{1}{2}} \,=\, r\,,\\
		\frac{a_o}{a_\infty} &\,=\, \left[\frac{1 \,+\, \left(\frac{\gamma-1}{2}\right)M_\infty^2}{1 \,+\, \left(\frac{\gamma-1}{2}\right) M_o^2}\right]^\frac{1}{2},\\
		K &\,=\, \begin{cases}
			\frac{1+2r}{(1+r)^2} &\text{for K-H mode},\\
			\frac{1}{1+r} &\text{for supersonic mode},\\
			\frac{1}{(1+r)^2} &\text{for subsonic mode}.
		\end{cases}
	\end{split}
\end{equation}

\begin{figure}
	\centering\includegraphics[width=0.75\linewidth]{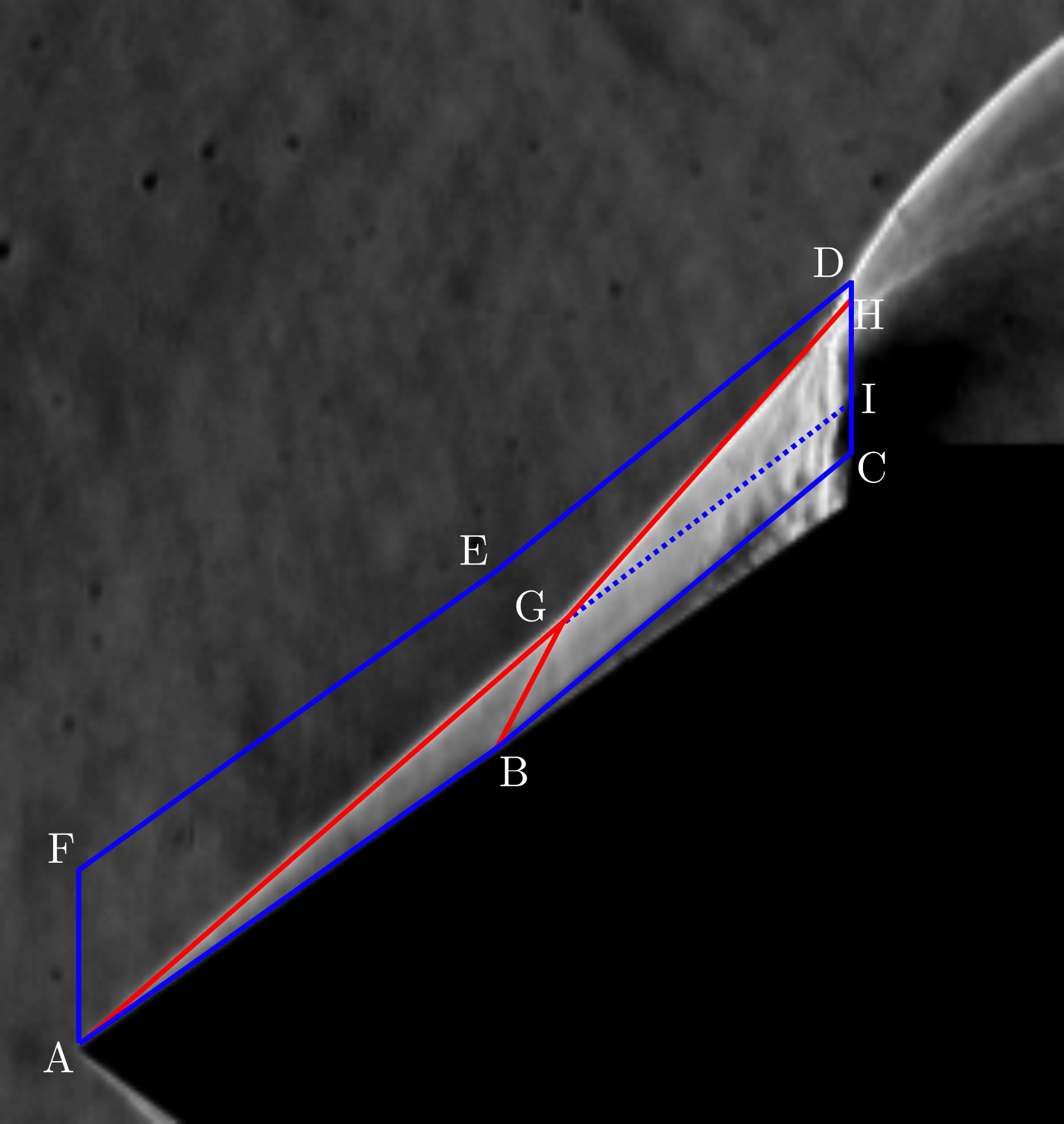}
	\caption{A representative CFD domain ABCDEF for the axisymmetric inviscid flow simulation over a a double cone model with $\theta_1 = 35^\circ$ and $\Lambda = 0.06$. Simulated Shock waves and shear layer are shown with solid red and dashed blue lines, respectively.}
	\label{fig:35deg_CFD}
\end{figure}

Using equations~\ref{eq:parameters} in equation~\ref{eq:dcrossiter}, the input parameters to calculate $St$ for a given double cone model are simply $M_\infty$, $M_o$ and $L_s/l_1$. In the present study, free-stream Mach number $M_\infty = 6$, and the non-dimensional shear layer length ($L_s/l_1$) is estimated from a time-averaged schlieren image for any particular experiment. An accurate estimate for $M_o$ however is not easy to obtain for all the oscillation cases using the schlieren data. For the anchored oscillation cases where the shear layer separates at the cone nose, the separation shock wave angle and the downstream flow properties can be calculated by considering flow deflection from an effective cone with half-angle ($\theta_1+\theta_s$) and solving Taylor-Maccoll equation \citep{anderson2016ebook}. Here, $\theta_s$ is the shear layer angle obtained from the time-averaged schlieren data. However, for free-oscillation cases, the downstream flow properties which depend on the successive flow deflections from the nose and separation shock waves cannot be calculated using approximations made in Taylor-Maccoll equation. Hence, to accurately obtain flow Mach number above the shear layer ($M_o$), two-dimensional axi-symmetric inviscid computational fluid dynamics (CFD) simulations were performed using a compressible flow solver in OpenFOAM \citep{kumar2022improved,kumar2021role}. These simulations are fairly inexpensive due to steady, inviscid and 2D nature of the flow simulations. A representative domain for a double cone model with $\theta_1 = 35^\circ$ and $\Lambda = 0.06$ is shown in figure~\ref{fig:35deg_CFD} (lines ABCDEF drawn in blue color). The slip boundaries AB and BC coincide with the fore-cone surface and observed shear layer in the time-averaged schlieren image, respectively. Free-stream conditions from the experiment are imposed at the boundary AF, and CDEF is considered as non-reflecting far-field boundary. The inviscid axisymmetric simulation with a slip boundary BC accounts for the flow turn at point B due to presence of shear layer in the experiment. The conical shock wave AG, separation shock wave BG, intermediate shock wave GH, a weak shock wave GI obtained from the numerical simulation are found to be in very close agreement with the experimental schlieren image. The Mach number at surface BC obtained from CFD simulations is considered as the average Mach number above the shear layer ($M_o$) for modeling purposes. The values of $L_s/l_1$, $M_o$, $a_o/a_i$ and $St$ for all the experimental cases are tabulated in table~\ref{tab:St_modeling}. From the table it is seen that for any given $\theta_1$, $M_o$ decreases with increasing $\Lambda$ as a consequence of increasing separation region size. Also, $M_o$ decreases with increasing $\theta_1$ (see cases with $\Lambda \approx 0.2$) as the conical shock wave angle increases at the cone nose. These observations are consistent with the earlier discussion in \S3.

\begin{table}
	\centering
	\caption{Measured values of non-dimensional shear layer length ($L_s/l_1$) and Strouhal number ($St$), estimated values of Mach number ($M_o$) and modeled values of Strouhal number based on three types of disturbances waves in the shear layer ($St_{\textup{KH}}$, $St_{\textup{sup}}$ and $St_{\textup{sub}}$) for all the considered experimental cases.}\label{tab:St_modeling}%
	\begin{tabular}{lllllllll}
	\hline
	$\theta_1$ & $\Lambda$ & $L_s/l_1$ & $M_o$& $a_o/a_i$ & $St$ & $St_{\textup{KH}}$ & $St_{\textup{sup}}$ & $St_{\textup{sub}}$\\ \hline
    8  &  0.13  &  1.037  &  4.08  &  0.518  &  0.186  &  0.209  &  \underline{0.186}  &  0.153 \\ \hline
8  &  0.15  &  1.046  &  3.88  &  0.538  &  0.191  &  0.205  &  \underline{0.183}  &  0.149 \\ \hline
8  &  0.17  &  1.053  &  3.723  &  0.555  &  0.191  &  0.203  &   \underline{0.18}  &  0.145 \\ \hline
8  &  0.19  &  1.061  &  3.618  &  0.566  &  0.192  &  0.201  &   \underline{0.178}  &  0.143 \\ \hline
8  &  0.21  &  1.067  &  3.53  &  0.576  &  0.191  &  0.199  &   \underline{0.175}  &  0.14 \\ \hline
10  &  0.08  &  0.818  &  4.4  &  0.488  &  0.3  &  0.267  &  0.24  &  0.2 \\ \hline
10  &  0.1  &  0.938  &  4.3  &  0.497  &  0.208  &  0.232  &   \underline{0.208}  &  0.173 \\ \hline
10  &  0.11  &  1.035  &  3.98  &  0.528  &  0.16  &  0.208  &  0.186  &  \underline{0.152} \\ \hline
10  &  0.13  &  1.04  &  3.887  &  0.537  &  0.18  &  0.207  &  \underline{0.184}  &  0.15 \\ \hline
10  &  0.15  &  1.049  &  3.717  &  0.555  &  0.181  &  0.204  &  \underline{0.181}  &  0.146 \\ \hline
10  &  0.17  &  1.061  &  3.548  &  0.574  &  0.186  &  0.2  &  \underline{0.177}  &  0.141 \\ \hline
10  &  0.19  &  1.071  &  3.389  &  0.593  &  0.183  &  0.197  &  \underline{0.173}  &  0.137 \\ \hline
10  &  0.21  &  1.088  &  3.358  &  0.597  &  0.184  &  0.193  &  \underline{0.17}  &  0.134 \\ \hline
10  &  0.25  &  1.095  &  3.088  &  0.632  &  0.183  &  \underline{0.189}  &  0.165  &  0.128 \\ \hline
15  &  0.08  &  0.957  &  3.7  &  0.557  &  0.174  &  0.223  &  0.198  &   \underline{0.16} \\ \hline
15  &  0.1  &  0.996  &  3.6  &  0.568  &  0.145  &  0.213  &  0.189  &   \underline{0.151} \\ \hline
15  &  0.12  &  1.042  &  3.4  &  0.592  &  0.15  &  0.202  &  0.178  &   \underline{0.141} \\ \hline
15  &  0.15  &  1.038  &  3.3  &  0.604  &  0.161  &  0.202  &   \underline{0.177}  &   \underline{0.14} \\ \hline
15  &  0.18  &  1.06  &  3.27  &  0.608  &  0.177  &  0.198  &   \underline{0.173}  &  0.136 \\ \hline
15  &  0.2  &  1.071  &  3.16  &  0.622  &  0.17  &  0.194  &   \underline{0.17}  &  0.133 \\ \hline
25  &  0.08  &  0.701  &  3.0  &  0.644  &  0.201  &  0.294  &  0.256  &   \underline{0.198} \\ \hline
25  &  0.11  &  0.777  &  2.9  &  0.658  &  0.171  &  0.264  &  0.229  &   \underline{0.175} \\ \hline
35  &  0.06  &  0.5  &  2.1  &  0.785  &  0.232  &  0.379  &  0.319  &   \underline{0.226} \\ \hline
	\end{tabular}
\end{table}

\begin{figure}
	\centering\includegraphics[width=\linewidth]{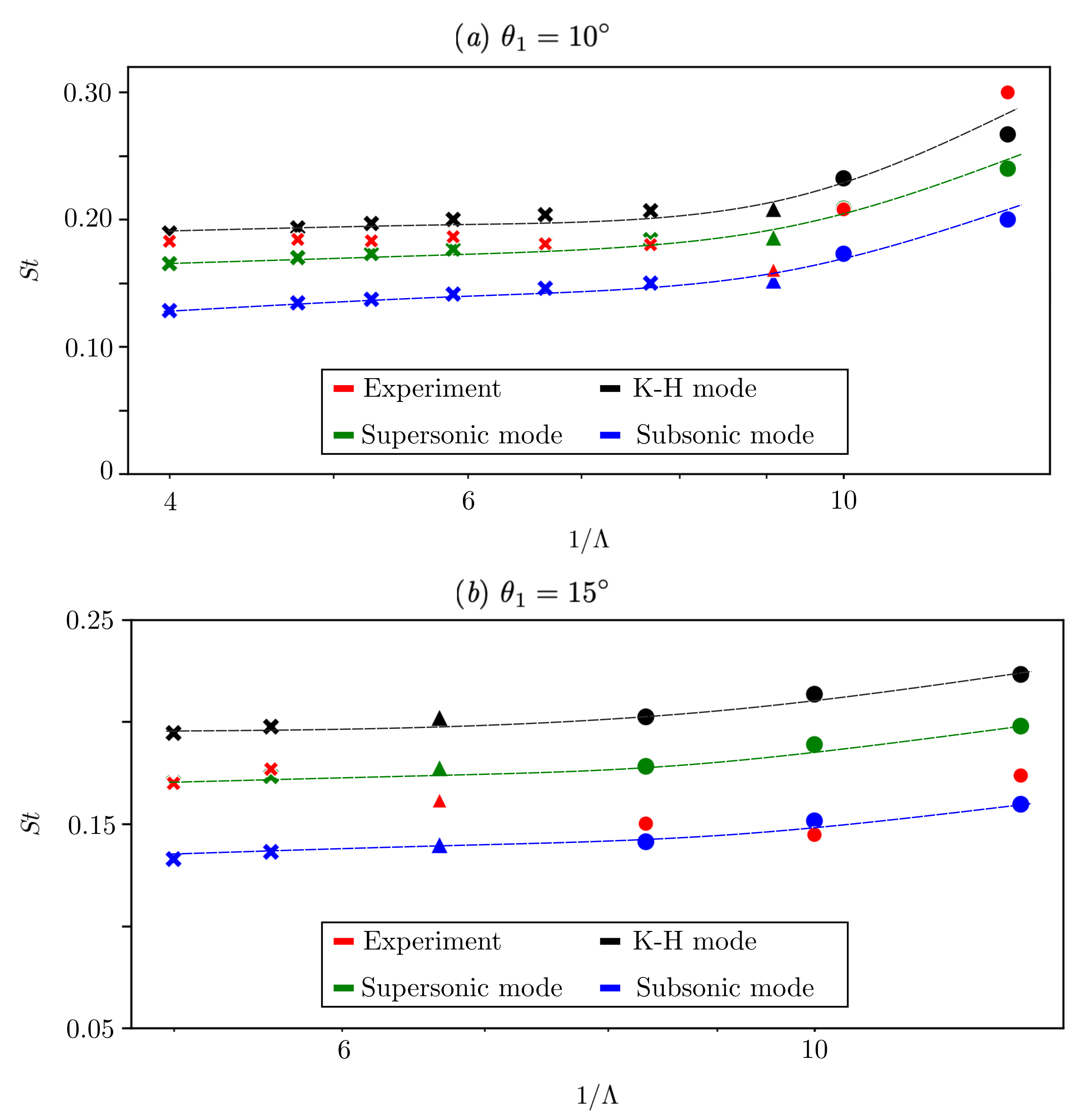}
	\caption{A comparison of predicted Strouhal number ($St$) from three instability types/modes of the shear layer (K-H mode, supersonic mode, subsonic mode) with experimentally observed Strouhal number for (\emph{a}) $\theta_1 = 10^\circ$ and (\emph{b}) $\theta_1 = 15^\circ$.  Marker shapes represent oscillations sub-types: free-oscillation ($\bullet$); anchored-oscillation ($\bold{\times}$); mixed-oscillation ($\blacktriangleup$).}
	\label{fig:St_prediction}
\end{figure}

Using the aeroacoustic model developed above (equations~\eqref{eq:dcrossiter} and ~\eqref{eq:parameters}), three values of Strouhal numbers, corresponding to three types of disturbance waves in the shear layer, are estimated for each combination of $\theta_1$ and $\Lambda$ studied experimentally. Figure~\ref{fig:St_prediction} shows a comparison of $St$ obtained from experiments and the model predictions for $\theta_1 = 10^\circ$ and $15^\circ$ as representative cases (comparisons for other values of $\theta_1$ can be found in table~\ref{tab:St_modeling}) that $St$ predictions from the model for anchored-oscillations are largely accurate when the supersonic mode speed is used for $u_c$. Though for $\theta_1 = 10^\circ$ and large values of $\Lambda$, it appears that the K-H mode speed gives a better prediction. For free-oscillations, the use of supersonic and subsonic mode speeds for $u_c$ gives a reasonably good prediction. These observations are consistent with the fact that for a zero thickness shear layer (the case of anchored-oscillations where the boundary layer separates at the cone tip with nominally zero boundary layer thickness), the subsonic disturbance mode is neutrally stable, thereby rendering the supersonic and K-H modes as the dominant modes that drive anchored-oscillations. For free-oscillations, a finite thickness of the shear layer promotes dominance of the subsonic mode, and thereby the oscillations are mostly driven by the same. The estimated values of $St$ for all the experiments in the present study are also tabulated in table~\ref{tab:St_modeling}, where the values closest to the experimentally observed $St$ are underlined.

Finally, we return to the observation made in figure~\ref{fig:Strouhalno} regarding the invariant behavior of $St$ for anchored-oscillations. From the model we know that two length scales come into play in determining $St$, one is the downstream propagation distance for shear layer disturbances and the other is the upstream propagation distance for acoustic disturbance in the separated region. For the case of anchored-oscillations, where the separation point is at the cone nose, which is also where the shear layer originates, both the relevant length scales are approximately equal to $l_1$, the slant length of the fore-cone. Since $l_1$ is the length scale used for scaling $f$ in equation~\eqref{eq:strouhal}, it is not entirely surprising that the $St$ does not show a strong dependence on $\theta_1$ or $\Lambda$. In case of free-oscillations however, the two relevant length scales depend on the location of the separation point $S$ on the fore-cone surface, and as seen from the experimental results in \S3, the location of $S$ depends on both $\theta_1$ or $\Lambda$. Hence, again, it is not entirely surprising that the $St$ for free-oscillations in figure~\ref{fig:Strouhalno} shows a clear dependence on $\theta_1$ or $\Lambda$. 

\section{Conclusions}\label{sec:Conclusion}
The oscillation state of flow unsteadiness for a double cone model was studied through extensive experiments at Mach 6. This study spans a wide range of the $\theta_1-\Lambda$ parameter space and identifies the influence of the governing geometric parameters on the qualitative and quantitative flow features. Two distinct sub-types of oscillations, namely free-oscillations and anchored-oscillations, were identified and the underlying oscillation dynamics was understood. The existence of two oscillation sub-types has not been reported in earlier literature. The global temporal scale of flow unsteadiness was extracted using spectral proper orthogonal decomposition, and the oscillation Strouhal number thus obtained is found to exhibit invariance for anchored-oscillations and a mixed scaling (with respect to $\theta_1$ or $\Lambda$) for free-oscillations. Coherent flow oscillations are understood to be sustained by an aeroacoustic feedback loop, and a simple oscillator model developed on the same basis performs well in predicting the experimentally observed oscillation Strouhal number. The model also provides helpful physical insight into the nature of the self-sustained flow oscillations over a double cone at high-speeds.

\section*{Acknowledgements}
The partial support for this research from SERB National Postdoctoral Fellowship (G.K. -- file no. PDF/ 2021/001358; V.S. -- file no. PDF/2020/001278), GoI Ministry of Education MTech Scholarship (A.G.K.), and Saroj Poddar Trust (S.D.) are gratefully acknowledged. The authors are thankful to B. M. Shiva Shankar, N. Shanta Kumar, and M. Harish for their assistance in the operations and maintenance of the hypersonic wind tunnel facility.

\bibliographystyle{jfm}
\bibliography{jfm-instructions}

\end{document}